%% file: 00-sample-sigplan.tex
\documentclass[sigplan,screen]{acmart}

\AtBeginDocument{%
  }

\usepackage{balance}
\usepackage[normalem]{ulem}
\usepackage{xspace}
\newcommand{\ignore}[1]{}
\newcommand{\DESIGN}{\mbox{T3}\xspace}
\usepackage{tikz}
\newcommand*\circled[1]{\tikz[baseline=(char.base)]{
            \node[shape=circle,draw,inner sep=0.5pt] (char) {#1};}}
\usepackage{listings}
\usepackage{color}
\usepackage{tikz}
\def\checkmark{\tikz\fill[scale=0.4](0,.35) -- (.25,0) -- (1,.7) -- (.25,.15) -- cycle;}
\usepackage{colortbl}
\definecolor{ForestGreen}{RGB}{34,139,34}
\usepackage{soul}
\usepackage{letltxmacro}
\LetLtxMacro{\oldst}{\st}                   
\makeatletter
\renewcommand\st[1]{\@bsphack\@esphack}%      %Turn off struck text when everything is done
\makeatother
\usepackage{caption}
\usepackage{graphicx}
\usepackage{breqn}
\usepackage{amsmath}
\usepackage{caption}
\usepackage{float}
\usepackage{tikz}
\usepackage{xurl}

\copyrightyear{2024}
\acmYear{2024}
\setcopyright{rightsretained}
\setcopyright{rightsretained}
\acmConference[ASPLOS '24]{29th ACM International Conference on
Architectural Support for Programming Languages and Operating Systems,
Volume 2}{April 27-May 1, 2024}{La Jolla, CA, USA}
\acmBooktitle{29th ACM International Conference on Architectural Support for
Programming Languages and Operating Systems, Volume 2 (ASPLOS '24), April
27-May 1, 2024, La Jolla, CA, USA}
\acmDOI{10.1145/3620665.3640410}
\acmISBN{979-8-4007-0385-0/24/04}

\begin{document}

\title{\DESIGN{}: \textit{T}ransparent \textit{T}racking \& \textit{T}riggering for Fine-grained Overlap of Compute \& Collectives}

\author{Suchita Pati}
\affiliation{%
  \institution{University of Wisconsin-Madison}
  \institution{Advanced Micro Devices, Inc.} 
  \country{United States}
  }
\email{spati@cs.wisc.edu}

\author{Shaizeen Aga}
\affiliation{%
  \institution{Advanced Micro Devices, Inc.}
  \country{United States}
}
\email{shaizeen.aga@amd.com}

\author{Mahzabeen Islam}
\affiliation{%
 \institution{Advanced Micro Devices, Inc.}
 \country{United States}
 }
 \email{mahzabeen.islam@amd.com}

\author{Nuwan Jayasena}
\affiliation{%
  \institution{Advanced Micro Devices, Inc.}
  \country{United States}
  }
\email{nuwan.jayasena@amd.com}

\author{Matthew D. Sinclair}
\affiliation{%
  \institution{University of Wisconsin-Madison}
  \institution{Advanced Micro Devices, Inc.}
  \country{United States}
  }
\email{sinclair@cs.wisc.edu}

\renewcommand{\shortauthors}{Pati et al.}

\input{001-abstract}

\begin{CCSXML}
<ccs2012>
   <concept>
       <concept_id>10010520.10010521.10010542.10010294</concept_id>
       <concept_desc>Computer systems organization~Neural networks</concept_desc>
       <concept_significance>300</concept_significance>
       </concept>
   <concept>
       <concept_id>10010147.10010169</concept_id>
       <concept_desc>Computing methodologies~Parallel computing methodologies</concept_desc>
       <concept_significance>300</concept_significance>
       </concept>
          <concept>
       <concept_id>10010520.10010521.10010528.10010534</concept_id>
       <concept_desc>Computer systems organization~Single instruction, multiple data</concept_desc>
       <concept_significance>300</concept_significance>
       </concept>
       <concept>
<concept_id>10010147.10010919</concept_id>
<concept_desc>Computing methodologies~Distributed computing methodologies</concept_desc>
<concept_significance>100</concept_significance>
</concept>
 </ccs2012>
\end{CCSXML}

\ccsdesc[300]{Computer systems organization~Single instruction, multiple data}
\ccsdesc[300]{Computer systems organization~Neural networks}
\ccsdesc[300]{Computing methodologies~Parallel computing methodologies}
\ccsdesc[300]{Computing methodologies~Distributed computing methodologies}

\keywords{Distributed Machine Learning, Collective Communication, Transformers, GPUs, Fusion, Fine-grained Overlap, Near-memory Computing}

\received{10 August 2023}
\received[revised]{3 January 2024}
\received[accepted]{8 January 2024}

\maketitle

\input{01-introduction}

\input{02-bkg}

\input{03-insight}

\input{05-proposal}

\input{06-methodology}

\input{07-evaluation}

\input{08-discussion}

\input{09-related}

\input{10-conclusion}
\input{11-ack}

\balance
\bibliographystyle{ACM-Reference-Format}
\bibliography{article}

\end{document}

%% file: 001-abstract.tex
\begin{abstract}

Large Language Models increasingly rely on distributed techniques for their training and inference. These techniques require communication across devices which can reduce scaling efficiency as the number of devices increases. While some distributed techniques can overlap, and thus, hide this communication with independent computations, techniques such as Tensor Parallelism (TP) inherently serialize communication with model execution. One approach to hide this serialized communication is to interleave it with the producer operation (of the communicated data) in a \textit{fine-grained} manner.
However, this fine-grained interleaving of communication and computation in software can be difficult.
Furthermore, as with any concurrent execution, it requires compute and memory resources to be shared between computation and communication, causing resource contention that reduces overlapping efficacy.

To overcome these challenges,
we propose \DESIGN{} which applies hardware-software co-design to transparently overlap serialized communication while minimizing resource contention with compute.
\DESIGN{} \textit{transparently fuses} producer operations with the subsequent communication via a simple configuration of the producer's output address space and requires minor software changes.
At the hardware level, \DESIGN{} adds a lightweight \textit{track} \textit{and trigger} mechanism to orchestrate the producer's compute, and communication. It further uses \textit{compute-enhanced memories} for communication's attendant compute.
As a result, \DESIGN{} reduces resource contention, and efficiently overlaps serialized communication with computation.
For important Transformer models like T-NLG, \DESIGN{} speeds up communication-heavy sublayers by 30\% geomean (max 47\%) and reduces data movement by 22\% geomean (max 36\%). Furthermore, \DESIGN{}'s benefits persist as models scale: geomean 29\% for sublayers in $\sim$500-billion parameter models, PALM and MT-NLG.

\end{abstract}

%% file: 01-introduction.tex
\section{Introduction}
\label{sec:intro}

\begin{figure}[tb!]
    \centering
    \includegraphics[width=\columnwidth, trim={0.5cm 12.5cm 19cm 4.5cm}]{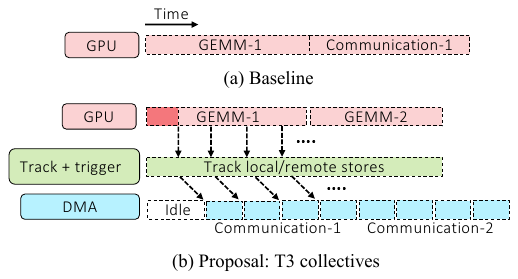}
    \vspace{-2ex}
    \caption{\DESIGN{} overview.}
    \label{fig:t3_overview}
    \vspace{-4ex}
\end{figure}

Deep neural networks (DNNs) have transformed society with significant accuracy improvements for tasks including speech recognition~\cite{toward-human-parity-conversational-speech-recognition}, image classification~\cite{Resnet, Krizhevsky12-alexnet, LinChen2013-nwInNW, SimonyanZisserman2014-imageRec, SzegedyLiu2015-deeperConv, SzegedyVanhoucke2015-cv}, machine translation~\cite{hassan2018achieving}, autonomous agents~\cite{LinZhang2018-avArch}, language processing~\cite{DevlinChang18-bert, RadfordWu2019-gpt2} and text generation~\cite{BrownMann2020-gpt3}.
This tremendous, transformative effect has been enabled by a virtuous synergy of (1) better hardware systems, (2) larger datasets, and (3) improved ML structures and algorithms that further benefit from more efficient hardware and larger datasets. This is especially true for Transformers, which have become popular for a wide range of tasks~\cite{StateofAI22} and have shown considerable strides in artificial general intelligence~\cite{DeepMind-Gato22}.
Transformers have had exponential growth in datasets and model size: parameters have increased from 340 million in BERT~\cite{DevlinChang18-bert} to 540 billion in PALM~\cite{chowdhery2022palm}.  Accordingly, their memory and computational demands have also increased, making them increasingly reliant on distributed techniques: multiple accelerators (e.g., GPUs) pooling their collective memory capacities and compute capabilities to collaboratively execute a DNN.
However, the resulting communication between devices has become a significant proportion of their execution time and has limited the scaling efficiency with increasing device count~\cite{klenk2020network,moolchandani2023amped,pati2023computation}.

\begin{figure*}[t!]
    \centering
    \includegraphics[width=0.90\linewidth, trim={3cm 10.5cm 5cm 5cm}]{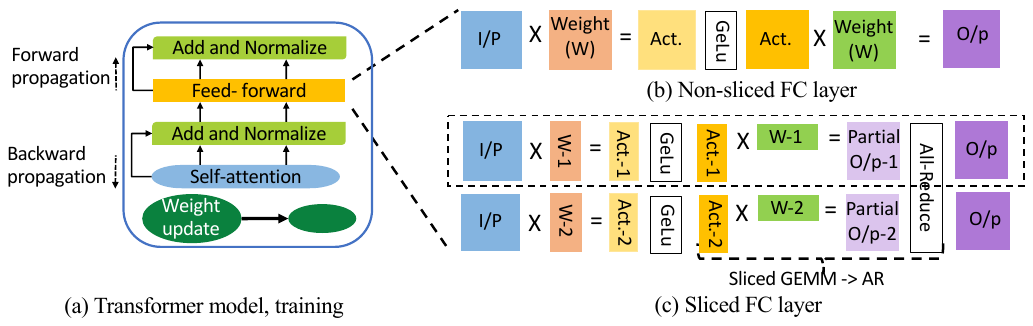}
    \caption{(a) Transformer (b) Fully-connected (FC) layer (c) Tensor-sliced FC layer with all-Reduce on the critical path.}
    \label{fig:bkg_transformer_tp}
\end{figure*}

Transformers frequently use two key distributed techniques in conjunction: data parallelism (DP) and model parallelism (MP).
DP parallelizes training by partitioning the dataset and replicating the model across devices, requiring communication and aggregation (\textit{all-reduce}) of gradients.
Conversely, MP partitions large models that cannot fit in a single device's memory.
Tensor-parallelism (TP), a type of MP, requires an all-reduce of layer outputs between devices as well.
Among these distributed techniques, TP's communication typically lies on the critical path of model execution, as shown in Figure~\ref{fig:t3_overview}(a) and can be a significant proportion of runtime ($\sim$45\%~\cite{pati2023computation}), resulting in a sub-linear increase in throughput as the number of devices increases. 

\noindent

While some prior works have sped up communication by up to 2$\times$ with \textit{in-network computation}, they are topology-dependent (requiring switches) and further, cannot eliminate serialized communication from the critical path~\cite{klenk2020network}.
Distributed techniques with abundant coarse-grained independent compute (e.g., DP) often overlap (and hide) communication with independent computations to improve efficiency.
Although serialized communication scenarios also offer such potential, they require a fine-grained overlap of computation and communication, which presents its own challenges.
Enabling their fine-grained overlap in current systems either requires expensive fine-grained synchronization~\cite{jangda2022breaking} or changes to matrix multiplication (GEMMs) kernels which can be disruptive to GPU software infrastructure~\cite{wang2022overlap} (Section~\ref{sec:chal_complex_sw}).
Furthermore, overlapped compute and communication contend for both compute units and memory bandwidth, reducing overlap's efficacy~\cite{jangda2022breaking,wang2022overlap} (Section~\ref{sec:chal-resource}).
Prior approaches that reduce contention only address coarse-grained overlap of compute and communication in cases like DP and lack support for fine-grained overlap in serialized collectives~\cite{rashidi2021enabling}.
Moreover, they rely on dedicated accelerators.
\textit{Therefore, no existing technique achieves a transparent overlap of serialized communication with computation while minimizing resource contention.}

To overcome these, we propose \DESIGN{} (Figure~\ref{fig:t3_overview}(b)).
\DESIGN{} \textit{transparently} fuses producer operations
with the subsequent communication by \textit{configuring the 
producer's output address space} to initiate communication 
directly on the producer's store, requiring minimal application
changes.
It uses a lightweight and programmable hardware
\textit{tracker} to track the producer/communication progress 
and \textit{triggers} communication using pre-programmed DMA commands, requiring no additional GPU compute resources for communication.
Furthermore, to reduce contention for memory bandwidth between the producer and communication, \DESIGN{} leverages recently proposed compute-enhanced memories~\cite{lee2021hardware,KimPark2021-gradPim}
to atomically update memory on stores, thus reducing memory 
traffic due to communication-related reductions.
Finally, \DESIGN{} employs a simple yet effective arbitration policy between the producer and communication memory streams to minimize any remaining contention.
Overall, \DESIGN{} transparently overlaps serialized
communication with minimal resource contention.
This improves compute and network utilization, and in turn,
can enable better throughput scaling with increasing device count.
We make the following key contributions: 

\begin{itemize}
 \item We propose \DESIGN{} which enables fine-grained overlap of serialized communication with its producer computation whilst lowering application impact and managing compute and memory interference.
 
 \item To manage application impact, \DESIGN{} configures the producer's output address space mapping to initiate communication on stores, requiring minor modifications to the producer kernels. 
 
 \item To manage compute resources contention, \DESIGN{} uses a lightweight programmable tracker that tracks producer progress and triggers communication using existing DMA engines requiring no additional compute resources.
 
 \item Finally, to tackle memory bandwidth contention between computation and communication, \DESIGN{} harnesses emerging near-memory compute technology to reduce data movement due to communication. Further, \DESIGN{} also devices a simple yet effective memory controller arbitration policy to better interleave computation and communication memory traffic. 

 \item Similar to prior work~\cite{KhairyNikiforov2020-ladm}, we extend Accel-Sim~\cite{KhairyShen2021-accelSim} to accurately model multi-GPU systems (6\% error). Our results show that \DESIGN{} speeds up sliced Transformer sub-layers from models like Mega-GPT-2~\cite{ShoeybiPatwary2019-megatronlm} and T-NLG~\cite{Microsoft2020-tnlg} by 30\% geomean (max 47\%) and reduces data movement by 22\% geomean (max 36\%). 
 Furthermore, \DESIGN{}'s benefits persist as models scale: geomean 29\% for sublayers in $\sim$500-billion parameter models, PALM and MT-NLG. 
 Overall, \DESIGN{} speeds up model training by up to 12\% and inference (prompt phase) by up to 15\%.

\end{itemize}

%% file: 02-bkg.tex
\section{Background \& Motivation}

\subsection{Transformers \& Need for Distributed Computing}
\label{subsec:bkg-transformers}

Transformers~\cite{VaswaniShazeer17-attention} have become the general-purpose architecture for a wide range of tasks/domains (e.g., for text, image)~\cite{StateofAI22}.
Models use the Transformer \textit{encoder} or \textit{decoder} as their basic building block, each with 
an attention sub-layer and a fully connected (FC) sub-layer (as shown in Figure~\ref{fig:bkg_transformer_tp}(a)) which manifest as matrix multiplication operations (GEMMs). Each layer also contains a few residual connections and layer normalizations which manifest as element-wise operations, and are often fused~\cite{ElHajjGomezLuna2016-klap, FousekFilipovivc2011-fuseGPUMap, SpringerWauligmann2017-fuseGPULang, WangLin2010-gpuKernelFusion} with the GEMMs. As shown in Figure~\ref{fig:bkg_transformer_tp}(b), these GEMMs entail multiplication of layers' weight matrices by an input matrix (with each vector representing an input token).
During training, the input matrices contain multiple tokens from one or more (if batched) input sequence(s).
During inference, there are two execution phases: a \textit{prompt} phase to process all tokens in the input sequence(s) and a \textit{token generation} phase to iteratively process and generate one token at a time for each input sequence~\cite{patel2023splitwise}.
The prompt phase operations are similar to those in training, while the generation phase has GEMMs with small input matrices or matrix-vector operations (GEMVs) if there is no batching.  

Most Transformer models' memory capacity requirements exceed a single device. Thus, they employ distributed techniques and use multiple accelerators (e.g., GPUs) collaboratively.
Furthermore, the aggregate computational capacity of multiple devices also accelerates training by enabling the processing of large input datasets in parallel.
Thus, since Transformers and their datasets (usually large corpora of unlabeled text) have increased by several orders of magnitude in size, distributed techniques are often mandatory and increasingly require many devices.
This scaling will only increase for future models.

\subsection{Distributed Techniques \& Associated Collectives}
Transformers employ many distributed techniques, each with associated \textit{communication} between devices.
Data parallelism (DP) trains model replicas on multiple devices, each on a  disjoint set of data, and requires a reduction of gradients every iteration. Tensor parallelism (TP)~\cite{ShoeybiPatwary2019-megatronlm} and pipeline parallelism (e.g., GPipe)~\cite{HuangCheng2019-gpipe} slice the model across multiple devices. While the former slices each layer requiring activation reduction, the latter partitions the model layer-wise requiring peer-to-peer transfer of activations. ZeRO-based optimizations~\cite{ZeRO-Infinity21} also slice model weights or offload them to slower but larger (e.g., CPU) memories, and require them to be gathered before layer executions. Finally expert parallelism~\cite{Kim21Expertllsm} partitions mixture-of-expert (MoE) models~\cite{rajbhandari2022deepspeed,Fedus21} such that each device hosts a single expert and requires exchange of input data based on input-to-expert mapping.
These communication patterns are handled by \textit{collectives} such as \textit{reduce-scatter}, \textit{all-reduce}, \textit{all-gather}, \textit{all-to-all}.
While most of this communication can be hidden by independent compute operations~\cite{moolchandani2023amped,PatiAga2021-demystifying,pati2023computation}, albeit with some resource contention~\cite{klenk2020network,rashidi2021enabling}, the all-reduce in TP is not (detailed in Section~\ref{subsec:bkg-tp-exposed}).
Thus, we focus on all-reduce in TP and discuss other techniques/collectives in Sections~\ref{subsec:disc-other-collectives} and~\ref{subsec:disc-distr-techniques}.

\subsection{All-Reduce \& Ring Implementations}
\label{subsec:bkg-reduce-scatter-ring}

\begin{figure}[t!]
  \centering
  \includegraphics[width=\columnwidth, trim={0.5cm 13.5cm 10.25cm 4.5cm}]{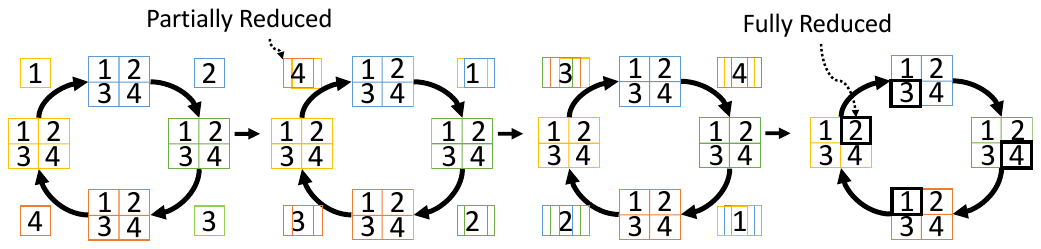}
  \caption{Ring implementation of reduce-scatter collective.}
  \label{fig:bkg-ring}
   \vspace{-1ex}
\end{figure}

The all-reduce (AR) collective reduces (element-wise sums) arrays from each of the devices.
Although there are multiple implementations of AR, one of the most bandwidth-efficient, and thus most commonly used, implementations is \textit{ring-AR}.
Ring-AR consists of a ring reduce-scatter (ring-RS) followed by a ring all-gather (ring-AG).
As shown in Figure~\ref{fig:bkg-ring}, ring-RS is done in multiple steps.
The arrays are chunked on each device, and during each step, all devices send their copy of a \textit{unique} chunk to their neighbor in the ring.
The devices then reduce their local copy of the chunk with the received copy and forward it to their neighbor in the next step.
With $N$ devices and the array chunked $N$ ways, this process requires $N-1$ steps until each device has a completely reduced copy of one chunk.
Ring-AG is similar but does not have reductions; it also requires $N-1$ steps until each device has all the reduced chunks.
In the remainder of the paper, we use AR, RS, and AG to refer to their ring implementations and discuss other implementations in Section~\ref{subsec:disc-other-collectives}.

\subsection{All-Reduce is on the Critical Path \& can be Large}
\label{subsec:bkg-tp-exposed}
\noindent
Transformers require tensor parallelism (TP)~\cite{ShoeybiPatwary2019-megatronlm} to increase the aggregate memory capacity available to them.
However, it requires ARs on the critical path (between layers).
Figures~\ref{fig:bkg_transformer_tp}(b) and ~\ref{fig:bkg_transformer_tp}(c) show the FC sub-layer's original operations versus the operations when sliced across two devices (TP=2 in Figure~\ref{fig:bkg_transformer_tp}(c)).
Each device (dotted box) only has a slice of the weights.
Since the GEMM corresponding to the second sliced weight only generates a partial output, it requires an AR before the next layer executes (highlighted by "Sliced GEMM$\rightarrow$AR").
These GEMM and AR operations execute as separate kernels and are serialized.

\begin{figure}[t!]
    \centering
    \includegraphics[width=\columnwidth, trim={1cm 11.7cm 15.75cm 4.5cm}]{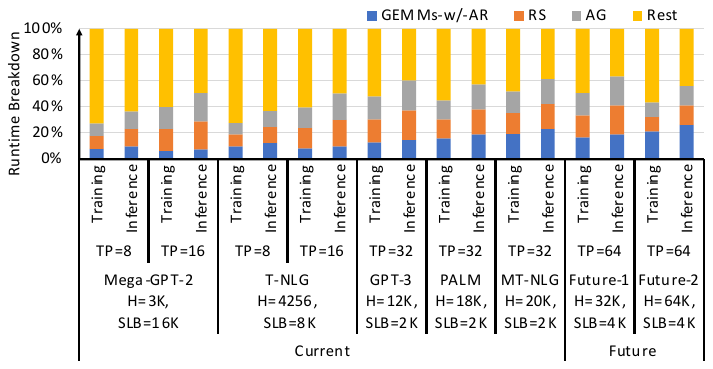}
    \vspace{-1ex}
    \caption{Transformer time spent on reduce-scatter (RS) and all-gather (AG) collectives as well as GEMMs which require collectives.}
    \label{fig:motiv-ar-percent}
   \vspace{-0.5ex}
\end{figure}

These serialized ARs can become a bottleneck.
Figure~\ref{fig:motiv-ar-percent} shows the execution time breakdown of Transformers between "Sliced GEMM$\rightarrow$AR" and other operations for multiple current and futuristic Transformers (setup detailed in Section~\ref{sec:eval-end-to-end},~\ref{sec:eval-app-setup}).
For large models (e.g., Mega-GPT-2, T-NLG) we consider 8- and 16-device TP.
For very large models (e.g., PALM, MT-NLG) we consider 32-way slicing, and for futuristic ones with one and ten trillion parameters, we consider 64-way sharding.
The increasing TP slicing is necessary because these models’ larger sizes cannot fit in 16 GPUs~\cite{pati2023computation}
and the increased slicing is also enabled by nodes with larger device counts~\cite{GH200, wang2022overlap}.
Like prior work~\cite{klenk2020network,moolchandani2023amped,pati2023computation}, we find that communication is a considerable fraction of the overall runtime: Megatron-GPT-2 (Mega-GPT-2) and T-NLG spend up to 34\% and 43\% of their training and inference (prompt phase) time on communication.
These trends also hold for the very large and futuristic Transformers: communication can be up to 46\% and 44\% of their runtime, respectively.
Additionally, since compute FLOPS scales much more than network bandwidth~\cite{comp_bw_scaling}, these proportions will only increase in the future.
For example, if the GEMMs become 2$\times$ faster, communication increases to 75\% of model execution time -- making scaling to multiple devices extremely inefficient and potentially leaving GPUs idle while communication happens.
Thus, addressing serialized AR is critical to Transformer scaling.

\subsection{Enabling Compute-Communication Overlap}
\label{sec:motiv_stages}
\noindent
Overlapping collective kernels with independent compute kernels has been key to scaling DNNs in other distributed approaches (e.g., DP, GPipe~\cite{HuangCheng2019-gpipe}).
While TP does not have independent kernels to overlap AR with, we observe that it can benefit from a \textit{fine-grained overlap} with the producer GEMM itself.
Transformer GEMMs have large outputs, which are tiled/blocked and require many GPU workgroups (WGs) to complete.
Consequently, a GEMM cannot always execute all its WGs concurrently on the limited number of GPU compute units (CUs).
Thus, a GEMM executes and generates output in multiple \textit{stages}, where each stage is a set of WGs that the CUs can accommodate.
This holds even for sliced GEMMs that require AR.
As shown in Figure~\ref{fig:insight_sliced_kdim}, GEMMs in TP are sliced in the $K$ (or dot-product) dimension which decreases compute per WG, but the output size, WG count, and WG stages remain the same.
We utilize this observation to enable fine-grained overlap: communication of one stage's output data can be overlapped with compute of the next stage.
However, achieving practical and efficient fine-grained overlap is challenging as we describe in Section~\ref{sec:challenges}.

%% file: 03-insight.tex
\section{Challenges With Fine-grained Compute-Communication Overlap }
\label{sec:challenges}

This section details key challenges with the fine-grained overlap of compute and communication.

\subsection{Complex \& Expensive to Implement in Software}
\label{sec:chal_complex_sw}
\noindent
The producer and collective operations execute as separate kernels on GPUs; the producer (GEMM) generates the data, after which the collective orchestrates their bulk communication and reduction.
Extending the software for their fine-grained interleaving
can be complex and expensive.
It would involve breaking the producer and collective into smaller kernels or using dynamic parallelism, both of which can increase launch overheads and synchronization costs.
Alternatively, it can be achieved by writing fused GEMM and collective kernels, but this can incur significant programming effort~\cite{ElHajjGomezLuna2016-klap, FousekFilipovivc2011-fuseGPUMap, SpringerWauligmann2017-fuseGPULang, WangLin2010-gpuKernelFusion}.
First, BLAS libraries have hundreds of GEMM kernels optimized for different input sizes and GPU architecture, generated via an expensive tuning process~\cite{Tensile}.
Second, collectives are also of different types, and each has implementations optimized for different topologies.
Creating fused kernels for every combination of GEMM and collective implementations can thus be extremely complex and expensive.
Hence, it is imperative to achieve a fine-grained overlap of compute and communication without altering GEMM implementations.

\begin{figure}[tb!]
    \centering
    \includegraphics[width=0.7\columnwidth, trim={2.5cm 12cm 13.5cm 2.5cm}]{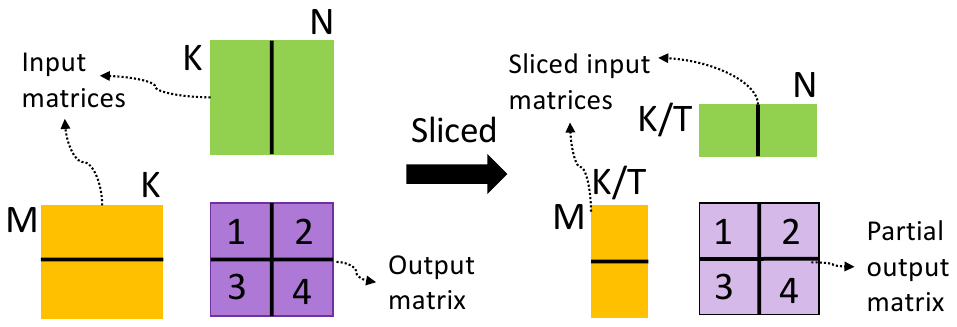}
     \vspace{-2ex}
    \caption{GEMM (left) when sliced in the dot-product dimension (right) still generates the same number of data blocks.}
    \label{fig:insight_sliced_kdim}
     \vspace{-0.5ex}
\end{figure}

\subsection{Resource Contention Between Producer \& Collective}
\label{sec:chal-resource}

Overlapped GEMM and AR contend for GPU resources, specifically compute units (CUs) and memory bandwidth, which slow down overall execution. 

\subsubsection{Compute Sharing}
\label{subsubsec:chal-resource-contention}

\begin{figure}[tb!]
    \centering
    \includegraphics[width=\columnwidth, trim={1cm 12cm 18cm 4.25cm}]{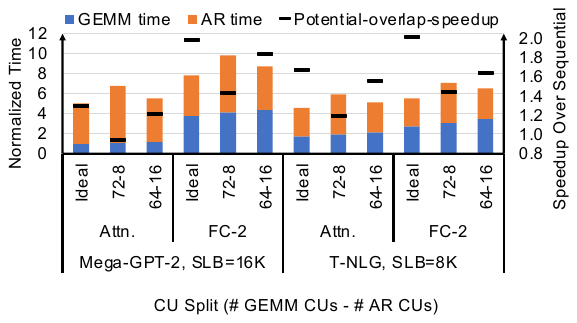}
    \caption{Evaluating how the benefits of overlapping GEMM and RS, across model layers, are impacted by compute unit (CU) sharing.  The X-axis shows how CUs are split between GEMM and AR, using the GPU setup from Table~\ref{tab:simulation-setup}, in the format $A$-$B$. $A$ represents the number of CUs the GEMM uses, while $B$ represents the number of CUs AR uses. Ideal assumes no sharing, the GEMM has all CUs, and AR is free.}
    \label{fig:chal_slowdown}
    \vspace{-3ex}
\end{figure}

Concurrently executing GEMM and AR kernels must share CUs and their components including L1 cache, LDS, and vector registers.
This contention may affect their performance relative to their isolated execution.
Figure~\ref{fig:chal_slowdown} evaluates the impact of concurrently executing GEMM and AR using our setup in Section~\ref{sec:multi-gpu-sim} and Table~\ref{tab:simulation-setup}.
Specifically, Figure~\ref{fig:chal_slowdown} shows the (normalized) GEMM and AR time for Mega-GPT-2 and T-NLG (with TP=8) sub-layers (Attn. and FC-2) when run in isolation with varying CU count splits (e.g., the 72-8 bars show GEMM's isolated execution time with 72 CUs and AR's with eight CUs).
For each case, it also shows \textit{potential-overlap-speedup}, the speedup overlapping AR and GEMM can obtain versus sequentially executing GEMM and AR when each has all 80 CUs.
We calculate the overlapped time as max(GEMM time, AR time).
The ideal case assumes no sharing impact: the GEMM has all the 80 CUs and the AR is fast but free (evaluated by running it with all 80 CUs in isolation).
As a result, the ideal case has the maximum potential overlap speedup of 1.67$\times$ geomean.
However, AR slows down considerably (geomean $\sim$41\% slowdown) when allocated only eight CUs (72-8 case) compared to when it had all CUs.
This significantly decreases the potential-overlap-speedup to 1.18$\times$ geomean.
While AR performance improves with 16 CUs (only $\sim$7\% slowdown in 64-16 case), GEMMs slow down (geomean $\sim$21\% slowdown) since they now only have 64 CUs.
Overall, while better than the 72-8 case, potential speedups fall short  (1.49$\times$ geomean) compared to the ideal case.
Moreover, this assumes no contention due to memory bandwidth sharing (discussed next) and thus underestimates slowdowns.
Overall, sharing of CUs reduces overlapping efficacy and it is crucial to preserve the compute resources dedicated to GEMMs.

\subsubsection{Memory Bandwidth Sharing}
\label{subsubsec:chal-resource-bw}

GEMM and AR kernels also compete for memory bandwidth when run concurrently.
As shown in Figure~\ref{fig:bkg-ring}, at each step AR kernels a) read an array chunk from memory to send it to one neighbor GPU and also b) write to memory the chunk it received from another neighbor.
Reduce-scatter (RS) additionally requires a memory read of the local chunk for reduction.
Moreover, the memory traffic due to AR communication can be bursty.
This additional, bursty memory traffic due to AR can slow down critical memory accesses by the producer GEMM, with the impact higher for GEMMs for which inputs do not fit in GPU's last level cache (LLC) as we will show in our evaluation in Section~\ref{sec:t3_speedups} and Figure~\ref{fig:t3_dram_contention}.
Thus, to enhance overlap efficiency, it is essential to limit memory traffic due to communication and/or limit their contention with GEMM. 

\begin{figure*}[tb!]
    \centering
    \includegraphics[width=0.9\linewidth, trim={0.5cm 11.5cm 10cm 4.5cm}]{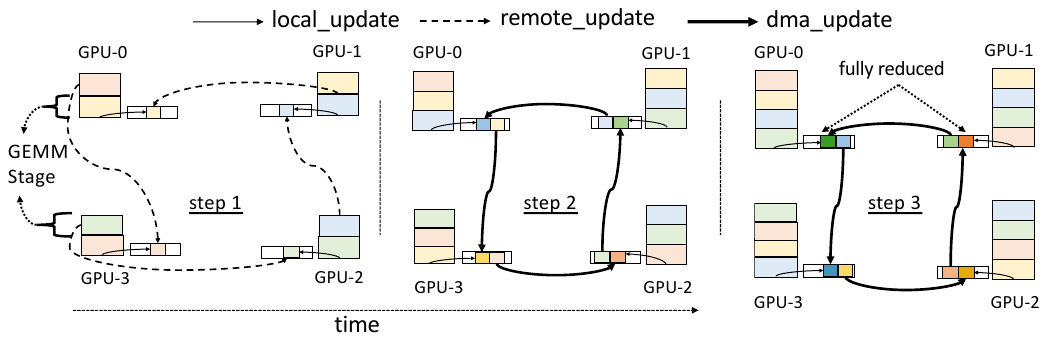}
    \caption{Overview of fused GEMM and ring reduce-scatter with \DESIGN{} on a four-GPU node.}
    \label{fig:fused_gemm_reduce_scatter}
\end{figure*}

Prior work also studied contention between communication and computation~\cite{rashidi2021enabling}, albeit in DP setups with coarse-grained GEMM and AR overlap.
They show that AR slows down by up to 2.4$\times$ when run concurrently with GEMMs, and the slowdown is even higher when run concurrently with memory-intensive embedding lookups in recommendation models.
For TP, they observe a 1.4$\times$ slowdown when executed concurrently with GEMMs.

%% file: 05-proposal.tex
\section{\DESIGN{}: Transparent Tracking \& Triggering}
\label{sec:fired_proposal}

To overcome the aforementioned challenges of complex software and resource contention with fine-grained overlap of compute and communication, we propose \DESIGN{}. 

\subsection{\DESIGN{} Overview}
\label{subsec:fired_overview}

Modern GPUs first execute the producer GEMMs and store their outputs in their local memory.
Afterwards they initiate the collective operation (Section~\ref{fig:bkg-ring}).
\DESIGN{} instead initiates the collective immediately as GEMMs generate data to enable fine-grained overlap. It uses a \textbf{\textit{track \& trigger}} mechanism to monitor GEMM's/collective's progress and to orchestrate communication, requiring no additional CUs (Section~\ref{subsec:t3_track_trigger}). It leverages \textit{\textbf{near-memory compute}} for reductions to reduce memory traffic due to communication (Section~\ref{subsubsec:t3_pim}).
Finally, it does these \textbf{\textit{transparently}}, with minor kernel modifications (Section~\ref{subsec:t3_addr_config}).

Figure~\ref{fig:fused_gemm_reduce_scatter} illustrates a four-device reduce-scatter (RS) overlapped with its producer GEMM.
This GEMM executes in multiple \textit{stages} of WGs dictated by its input and kernel implementation (Section~\ref{sec:motiv_stages}), while RS executes in multiple \textit{steps} dictated by the number of devices involved (Section~\ref{subsec:bkg-reduce-scatter-ring}).
For simplicity of illustration, we show the number of GEMM stages to be one more than the number of required ring steps.
In each step, a GEMM stage's execution and reduction of its output happen in parallel to the communication of the previous stage output. In the first step, the output is communicated to remote devices directly by the GEMM (\textit{remote\_update}).
The later, \textit{steady state}, steps require a DMA (\textit{dma\_update}).
For $N$ devices, this steady state step is performed $N-2$ times, on different chunks.
Focusing on GPU-0 in the steady state, step-2, as shown in Figures~\ref{fig:fused_gemm_reduce_scatter}, the GPU executes/generates output for GEMM stage-3 while also receiving (via DMA) a copy of stage'3 output (blue) from its neighbor, GPU-1.
This occurs in parallel to GPU-0's DMA of the reduced copy of GEMM stage-2 data (yellow) to GPU-3, thus overlapping communication.  
\DESIGN{} leverages \textit{near-memory computing} (NMC) to atomically update memory locations on these local and DMA updates, resulting in a partially reduced copy of the stage-3 chunk without requiring additional reads or GPU CUs (Section~\ref{subsubsec:t3_pim}). Once they complete,
GPU-0 initiates a \textit{dma\_update} of the chunk to its neighbor's (GPU-3) memory as shown in step-3. This automatic tracking of updates and DMA triggering is done using a lightweight and programmable hardware Tracker, further reducing dependency on GPU CUs (Section~\ref{subsec:t3_track_trigger}). 
These remote / DMA updates are done transparently by configuring the GEMM's output address mapping, with minor application and kernel modifications (Section~\ref{subsec:t3_addr_config}).

We also make minor runtime and hardware changes to improve \DESIGN{}'s performance.
To enable the perfect overlap of GEMM and RS in Figure~\ref{fig:fused_gemm_reduce_scatter}, we stagger the scheduling of GEMM workgroups (WGs) across GPUs (Section~\ref{subsec:t3_addr_config}).
Moreover, we also augment the memory system with a simple yet effective memory controller arbitration (MCA) policy to manage memory contention between compute and communication (Section~\ref{subsec:t3_mca}).

\begin{figure}[tb!]
    \centering
    \includegraphics[width=0.5\columnwidth, trim={1cm 10.5cm 22cm 4.5cm}]{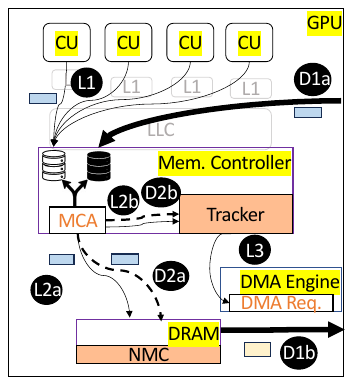}
    \caption{GPU with highlighted \DESIGN{} enhancements (in orange) executing a steady-state fused GEMM-RS step.}
    \label{fig:t3_system}
\end{figure}

Figure~\ref{fig:t3_system} shows a GPU with \DESIGN{}'s enhancements (in orange) executing the steady state step described above.
The GPU executes the GEMM to generate local updates for a stage (\circled{L1}). Concurrently the GPU receives DMA updates for the same stage (\circled{D1a}) and sends DMA updates for the previous stage (\circled{D1b}). At the memory controller, the modified \textit{MCA} arbitrates between the local and DMA traffic to prevent contention. Following this, the updates are sent to \textit{NMC-enhanced DRAM} (\circled{L2a},\circled{D2a}) while the \textit{Tracker} is updated with their progress (\circled{L2b},\circled{D2b}). Once the Tracker observes the required local and DMA updates to a memory region, it triggers their DMA transfer to the neighbor GPU (\circled{L3}).

We use the 4-GPU GEMM-RS overlap as a running example to describe \DESIGN{}. RS is more challenging to overlap due to reductions and extra memory traffic. Further, the ring configuration is more complex than others.
Thus, we detail \DESIGN{} using ring-RS and discuss additional collectives in Section~\ref{subsec:disc-other-collectives}.

\subsection{\textbf{\DESIGN{} Tracking \& Triggering}}
\label{subsec:t3_track_trigger}

\begin{figure*}[t!]
    \centering
    \includegraphics[width=\linewidth, trim={1.75cm 11.75cm 4.25cm 4.25cm}]{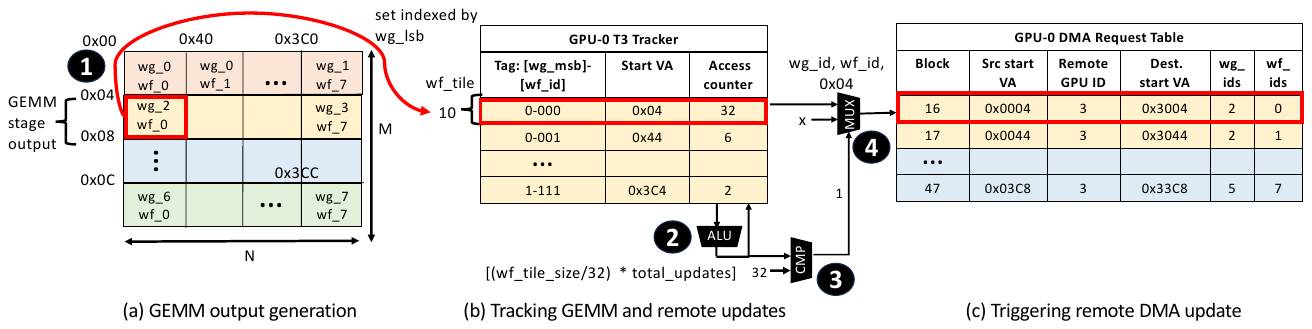}
    \caption{\DESIGN{} Track \& Trigger.}
    \label{fig:t3_track_trigger}
    \vspace{-0.5ex}
\end{figure*}

\DESIGN{}'s \textit{programmable track \& trigger} mechanism is key to transparently enabling fine-grained overlap of producer and collective without using compute resources.
As shown in Figure~\ref{fig:t3_track_trigger}, \DESIGN{} automatically transfers copies of data between devices when ready (e.g., in Figure~\ref{fig:fused_gemm_reduce_scatter}, \DESIGN{} triggers DMA update of stage-2 data from GPU-0 to GPU-3 once both GPU-0's local and GPU-1's remote updates are complete).
This is enabled by a lightweight \textit{Tracker} at the memory controller, that tracks local and remote/DMA accesses to memory regions and triggers a DMA transfer once the required accesses are complete.
Since the condition when a DMA is triggered (e.g., number of remote and local updates) and DMA transfer details (e.g., addresses, operation type) vary per collective type and implementation, they are programmed ahead of time using address space configuration (detailed in Section~\ref{subsec:t3_addr_config} and Figure~\ref{fig:t3_ring_addr_map_config}).

\subsubsection{Tracker}
\label{sec:t3_tracker}

The Tracker tracks both local and remote memory updates of a GEMM stage and triggers its DMA.
As shown in Figure~\ref{fig:t3_track_trigger}(a) and (b), it does so at wavefront (WF, i.e., a group of threads that execute in lockstep) granularity~\circled{1} -- i.e., the Tracker tracks the memory region a WF updates.
This assumes tiled GEMM implementations and that each WF/WG generates a complete tile of data, as is the case in all evaluated GEMMs~\cite{cutlass_nvidia,cublas}.
However, \DESIGN{} can also handle other implementation  (Section~\ref{subsec:disc-other-impl}).
An update increments the counter at its corresponding WF's ($wf\_id$) Tracker entry~\circled{2}.
This is done by all local, remote, and DMA updates that arrive at the GPU's memory controller (e.g., GPU-0 does not track GEMM stage-1 as its WFs neither write locally nor are its remote updates received).
The incremented counter value is checked for a maximum threshold, which is set to the product of WF output size ($wf\_tile\_size$) and the total updates expected per element~\circled{3}.
The $wf\_tile\_size$ is determined by the GPU driver 
using the output size and WF count ($(M*N)/\#WF$). 
The total updates expected per element for ring-RS is two but changes with collective type/implementation and is thus configurable (detailed in Section~\ref{subsec:t3_addr_config}).
Once the threshold is reached, 
the final write triggers the DMA (~\circled{4} in Figure~\ref{fig:t3_track_trigger}(c) and detailed in Section~\ref{sec:t3_dma}).
The Tracker is checked once the accesses are enqueued in the memory controller queue (MCQ) and thus are not in the critical path.

WF-based tracking is beneficial as a producer's (or GEMM's) stage may not update contiguous memory regions.
As shown in Figure~\ref{fig:t3_track_trigger}(a) this can happen due to column-major allocation of arrays in BLAS libraries~\cite{rocblas,cublas} and row-major scheduling.
This makes address-based tracking expensive (requires storing several addresses or complex table indexing functions) which WF-based tracking avoids.
The Tracker has a total of 256 entries, indexed using the workgroup (WG) ID's LSBs, $wg\_lsb$ (8 bits).
Each entry is set associative and is tagged using $wg\_msb, wf\_id$. 
$wg\_msb$ is $log\_2(max WGs per stage/256)$ bits and $wf\_id$ is three bits for a maximum of eight WFs per WG.
We set the maximum entries based on the maximum WGs possible in a producer stage. 
Each entry has a starting virtual address (smallest address per WF), and an accesses counter, making the Tracker size 19KB.
The tracking additionally requires the source $wg\_id$ and $wf\_id$ as metadata in memory accesses and forwarding of their virtual addresses to the memory controller (to trigger the DMA in Section~\ref{sec:t3_dma}).

\subsubsection{Triggering DMA}
\label{sec:t3_dma}

Once the required accesses to a WF's memory region are issued, \DESIGN{} DMAs the data to the remote GPU (~\circled{4} in Figure~\ref{fig:t3_track_trigger}(c)). As shown in Figure~\ref{fig:t3_track_trigger}(c), the DMA commands are pre-programmed by the GPU driver and are configurable (detailed in Section~\ref{subsec:t3_addr_config} and Figure~\ref{fig:t3_ring_addr_map_config}) as the DMA regions/operations can differ based on the collective type and implementation.
The granularity of the DMA block/table entry is set to be equal to or larger than the Tracker granularity ($wf\_tile$).
The memory access which completes the required accesses at the Tracker entry (Section~\ref{sec:t3_tracker}) marks the corresponding DMA entry ready and also populates it with the $wg\_id$ and $wf\_id$ which are required by the destination GPU's Tracker.
If DMA blocks are a multiple of $wf\_tile$, an additional counter per DMA entry can track their completion.
Using the pre-programmed starting source/destination virtual address, $wf\_tile\_size$, and the output dimensions (M, N),
the DMA engine dynamically generates the remaining virtual addresses to initiate the DMA.

\subsection{\textbf{Near-Memory Reductions}}
\label{subsubsec:t3_pim}

To perform reductions on producer and DMA updates without occupying 
GPU compute resources, \DESIGN{} leverages compute-enhanced memories.
We assume an HBM-based DRAM architecture with near-memory op-and-store support as has been proposed by recent works~\cite{nai2017graphpim,pawlowski2011hybrid}.
We envision such compute support to be implemented via ALUs near DRAM banks as has recently been proposed by memory vendors~\cite{KimPark2021-gradPim,lee2021hardware}.
However, \DESIGN{} can also leverage other reduction substrates (Section~\ref{subsec:disc-other-substrate}).

\DESIGN{} leverages this near-memory computing (NMC) capability to enable GEMM stores and DMA transfers to directly update and reduce copies of data, when required by collectives.
For DMA transfers, the operation type (store vs. updates) is directly specified in the command (address space configuration in Figure~\ref{fig:t3_ring_addr_map_config} and Section~\ref{subsec:t3_addr_config}).
For GEMMs, we utilize two flags.
First, we use an "uncached" flag during memory allocation to ensure that the output is not cached in any GPU's caches (such allocations are supported in existing GPUs).
Thus, writes are directly sent to DRAM which acts as the point of aggregation for all (local, remote, DMA) updates.
The queuing of updates in the memory controller queue guarantees their atomicity; at any given time, only a single instruction can be issued and executed by near-bank ALUs. 
Second, we use an "update" flag in the GEMM API call to enable stores of the GEMM to update the DRAM.
The "update" flag is sent (via kernel packets~\cite{amd-kernel-object}) to the CUs to tag the kernel's stores with one-bit "update" info (similar to prior work~\cite{jog2014application,jog2016exploiting,pattnaik2019opportunistic}). These are processed by the memory controller to generate the op-and-store commands.

\begin{figure}[tb!]
    \centering
    \includegraphics[width=\columnwidth, trim={0.5cm 14.5cm 17.2cm 4.5cm}]{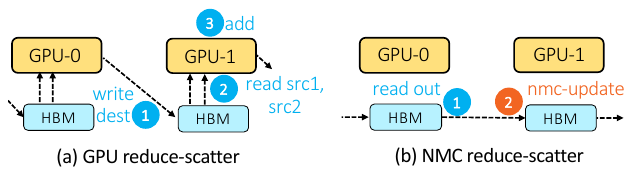}
    \caption{HBM reads \& writes in steady-state reduce-scatter step.}
    \label{fig:t3_nmc}
\end{figure}

In addition to freeing up CUs for GEMMs, NMC helps reduce memory traffic due to communication. Figure~\ref{fig:t3_nmc} shows memory accesses in a steady-state RS step in baseline and with \DESIGN{}.
In baseline RS, CUs read two copies of data (local copy, and received copy from the previous neighbor)
and write the reduced data to the next neighbor's memory.
\DESIGN{} only requires one read of the data to DMA update the neighbor GPU memory using NMC. Overall, \DESIGN{} with NMC reduces the dependence on GPU CUs and further reduces (or eliminates, \textit{direct-RS} in Section~\ref{subsec:disc-other-collectives}) data movement required for communication. 

\subsection{\textbf{Configuring Producer's Output Address Space}}
\label{subsec:t3_addr_config}

\begin{figure}[tb!]
    \centering
    \includegraphics[width=0.7\columnwidth, trim={1cm 12cm 19cm 5.5cm}]{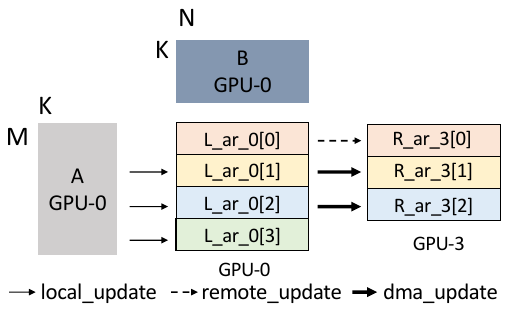}
    \caption{Remote address mapping for \DESIGN{} GEMM-RS over four GPUs.}
    \vspace{-2ex}
    \label{fig:t3_addr_space}
\end{figure}

\begin{figure}[tb!]
    \centering
    \includegraphics[width=\columnwidth, trim={0.5cm 11.5cm 17cm 4.5cm}]{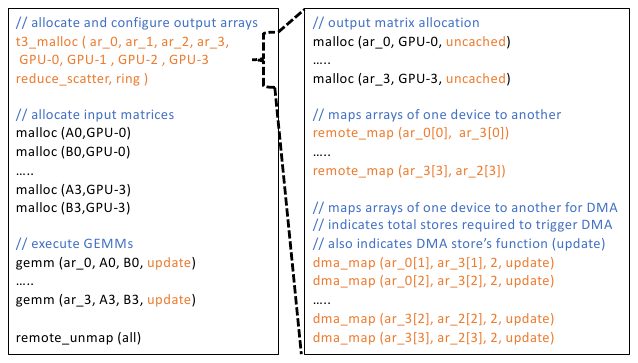}
    \caption{Configuring producer output for \DESIGN{} GEMM-RS over four GPUs.}
    \label{fig:t3_ring_addr_map_config}
    \vspace{-3.5ex}
\end{figure}

Modifying producer kernels, especially for many GEMMs of different shapes and sizes, to fuse and overlap collectives, can be impractical (Section~\ref{sec:chal_complex_sw}).
\DESIGN{} avoids this by configuring the producer's output address space mapping which is used to program the Tracker and DMA commands.
Figures~\ref{fig:t3_addr_space} and~\ref{fig:t3_ring_addr_map_config} show this 
configuration for GPU-0 from the fused GEMM-RS example in   Figure~\ref{fig:fused_gemm_reduce_scatter}.

Since there are four devices, GEMM's output array is chunked four ways.
In GPU-0, the GEMM writes its stage-1 output directly to GPU-3's memory (step-1 in Figure~\ref{fig:fused_gemm_reduce_scatter}), while its stage-2 and stage-3 output is first written to local memory and later DMA'd to GPU-3 (stage-4 is only written locally once and is not DMA'd).
Thus, GPU-0 requires memory mappings of these chunks with that of GPU-3 as shown in Figure~\ref{fig:t3_addr_space}.  
This configuration differs per collective type and topology-optimized implementation (see Section~\ref{subsec:disc-other-collectives}) and, similar to modern collective implementations, can be pre-defined in collective libraries~\cite{rccl,nccl}.
Figure~\ref{fig:t3_ring_addr_map_config} shows an example of this using pseudo-code.

The configuration in Figure~\ref{fig:t3_ring_addr_map_config} defines this mapping for the GEMM output using two different API calls: \textit{remote\_map} and \textit{dma\_map}.
\textit{remote\_map} is used for fine-grained remote writes/updates (for stage-1), which uses existing GPU support for peer-to-peer load/store by threads~\cite{gpudirect}.
Conversely, \textit{dma\_map} is used for coarse-grained DMA writes/updates (for stage-2,3) which leverages existing support 
for memory copies by DMA engines in GPUs (DirectGMA and others~\cite{gpudirect,muthukrishnan2021efficient,muthukrishnan2021gps}).
A \textit{dma\_map} call also defines the DMA functionality (store vs. update), and its triggering condition (number of stores/updates per element).
It can also be extended to specify granularity ($wf\_tile$s per DMA block in Figure~\ref{fig:t3_track_trigger}(c)).
These calls are used to pre-program the Tracker and DMA commands to enable automatic communication of data when ready (Section~\ref{subsec:t3_track_trigger}).

Fusion in ring-based collectives also benefits from producers (on different devices) generating data chunks in a staggered manner.
In Figure~\ref{fig:fused_gemm_reduce_scatter}, GPUs stagger the generated data by one stage; in step-1, GPU-0 executes stage-1, while GPU-1 executes stage-2, and so forth. This is enabled by staggering WG scheduling across devices. Alternatively, it can also be enabled by fetching appropriate implementation from BLAS libraries with staggered output tile-to-WG mapping amongst producer kernels.
Overall, configuring the output address space mitigates  
the need to change GEMM implementations to enable fusion with collectives.

\subsection{Communication-aware MC Arbitration (MCA):}
\label{subsec:t3_mca}

Finally, careful scheduling of memory accesses by the producer kernel and those resulting from communication
is crucial to efficiently overlap them.
In Section~\ref{sec:t3_layer_speedups} we show that a memory controller (MC) arbitration policy which a) round-robins between issuing memory accesses from the compute and communication streams and b) falls back to the other stream if the current stream is empty, results in producer kernel slowdowns.
Communication-related memory accesses appear in bursts and can occupy DRAM queues, stalling the compute kernel's critical memory reads/writes.
Simply prioritizing producer kernel accesses as they appear is also insufficient as prior communication-related memory accesses may already occupy DRAM queues. Finally, giving the local compute stream dedicated access results in wasted cycles and memory bandwidth underutilization.
Thus, an efficient overlap of compute and communication requires a dynamic arbitration policy that addresses both contention and under-utilization.

We implement a simple yet dynamic arbitration policy to overcome this.
The MC always prioritizes compute stream accesses, but if empty, falls back to communication stream. 
Additionally, it monitors the DRAM queue occupancy and only issues communication-related accesses if occupancy is below a threshold.
This ensures sufficient
room in the queues for future compute stream accesses and prevents their stalls.
The occupancy threshold depends on the memory-intensiveness of compute kernels (e.g., smaller if memory-intensive, and vice-versa).
This is determined dynamically: MC detects the memory intensiveness of a kernel by monitoring occupancy during its isolated execution (the first stage in Figure~\ref{fig:fused_gemm_reduce_scatter}).
Finally, the MC tracks cycles elapsed since the last issue from the communication stream and prioritizes it if it exceeds a limit to ensure it is not starved.
Additionally, the communication stream is drained at the producer kernel boundary.    

%% file: 06-methodology.tex
\section{Methodology}
\label{sec:method}

\subsection{Setup}
\label{sec:method-setup}

\subsubsection{Multi-GPU Simulation}
\label{sec:multi-gpu-sim}

Although a number of popular GPU simulators are publicly available~\cite{BaoSun2023-naviSim, GutierrezBeckmann2018-gem5GPU, lew19, RoartySinclair2020-gem5GPU}, we chose to evaluate \DESIGN{} using Accel-Sim~\cite{KhairyShen2021-accelSim} because it provides high fidelity for modern GPUs~\cite{KhairyJain2018-voltaGPGPUSim}.
Like prior work~\cite{KhairyNikiforov2020-ladm}, we extended Accel-Sim to simulate a multi-GPU system.
We observe that in a multi-GPU DNN setup all GPU's executions are homogeneous (Figures~\ref{fig:bkg_transformer_tp} and \ref{fig:t3_nmc}).
Thus, we evaluate both our multi-GPU baseline and \DESIGN{} by modeling all the activities pertaining to a single GPU.
This includes modeling the Tracker which is accessed/updated in parallel with the store/DMA operations and uncached NMC updates.
Although we do not model the DMA engine in the simulator, we do model its inter-GPU communication (communication resulting from RS both in the baseline and \DESIGN{}'s fused GEMM-RS) by executing the compute operations (e.g., GEMM~\cite{cutlass_nvidia,cublas}) in Accel-Sim and using Accel-Sim's front-end tracing functionality to inject the additional inter-GPU communication traffic.
The Tracker’s DMA triggering overheads are negligible since the DMA commands are pre-queued during the setup process (Figure~\ref{fig:t3_ring_addr_map_config}) as is often done, especially for ML operations which are repetitive~\cite{hwang2023ark}.
Table~\ref{tab:simulation-setup} details the GPU configuration we use to evaluate \DESIGN{}, which is the latest GPU architecture Accel-Sim completely supports.
Commercial GPUs with such a configuration support a 150 GB/s interconnection ring bandwidth~\cite{dgx-v100}.
Since recent GPUs frequently scale compute faster than other resources, we also evaluate another configuration with increased CU count while the other parameters stay the same in Section~\ref{sec:eval_hardware_scale}.

\begin{table}[tb]
  \centering
  \resizebox{0.8\columnwidth}{!}{%
  \input{tables/sim-setup}
  }
  \caption{Simulation setup.}
  \vspace{-3ex}
  \label{tab:simulation-setup}
\end{table}

\begin{figure}[t!]
    \centering
    \includegraphics[width=\columnwidth,trim={5.5cm 13.5cm 13.5cm 6cm},clip]{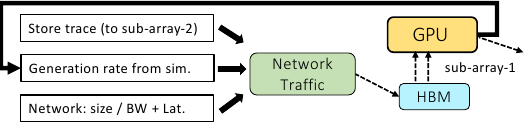}
    \vspace{-4ex}
    \caption{Simulating multi-GPU reduce-scatter.}
    \label{fig:multi-gpu-setup}
\end{figure}

\begin{figure}[t!]
    \centering
     \includegraphics[width=\columnwidth,trim={0cm 11.2cm 19cm 4.75cm},clip]{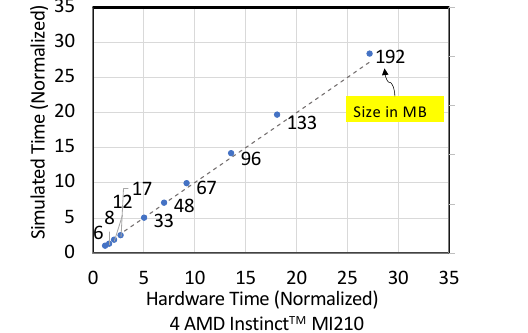}
    \caption{Validation of multi-GPU reduce-scatter simulation.}
    \vspace{-3ex}
    \label{fig:validation}
\end{figure}

Figure~\ref{fig:multi-gpu-setup} describes our multi-GPU simulation of RS.
In each RS step, a GPU performs a reduction of a sub-array and sends it to the neighbor GPU while also receiving a
reduced sub-array (corresponding to a different chunk) from another neighbor GPU (Figures~\ref{fig:bkg-ring} and \ref{fig:t3_nmc}(a)).
The simulator executes the reduction of the array as-is.
Simulating the incoming network traffic requires: (a) determining packet addresses, (b) generating packets at the appropriate rate, and (c) modeling the interconnect costs.
Packet addresses are determined using the store trace of WGs from the reduction kernel.
Next, since GPU executions are homogeneous, remote traffic is
generated at the same rate as the GPU generates the reduction output (which is filtered out to be sent to remote GPU).
This also implicitly includes slowdowns due to compute/communication interference at the remote GPU.
Finally, we add the interconnect costs to these packets as they arrive, assuming a simple link bandwidth and latency model of the interconnect.
To validate this setup, we compare simulated RS times on four GPUs with hardware measurements from a four-GPU node with AMD Instinct\texttrademark\ MI210 GPUs~\cite{MI210} with same ring network bandwidth as simulated (Table~\ref{tab:simulation-setup}).
Figure~\ref{fig:validation} shows that simulation closely follows hardware trends for a range of sizes (6-192 MB): 6\% geomean error versus the ideal dotted line.

\noindent
\textbf{Near-Memory Computing}:
We modify the simulator's HBM to model NMC updates.
Further, memory vendor proposals indicate that NMC operations can be issued without a significant increase in DRAM timings; back-to-back NMC operations can be issued to the same bank group with the same column-to-column access (CCDL) delay~\cite{KimPark2021-gradPim}. To model the additional cost of NMC op-and-store operations (Section~\ref{subsubsec:t3_pim}), we modify the simulator's HBM to use a 2$\times$ higher CCDL delay (termed CCDWL) following those operations (see Table~\ref{tab:simulation-setup}).

\noindent
\subsubsection{End-to-End Transformer Iteration}
\label{sec:eval-end-to-end}

To evaluate end-to-end iterations with \DESIGN{}, we scale the GEMMs and RS times in the baseline Transformer breakdown (shown in Figure~\ref{fig:motiv-ar-percent}) by their simulated speedups (described in Section~\ref{sec:multi-gpu-sim}).
We leverage a combination of hardware data and analytical modeling as done by prior works~\cite{pati2023computation,moolchandani2023amped} to get the end-to-end breakdowns of models in their distributed setups.
We use a single-GPU mixed-precision~\cite{micikevicius2018mixed} execution of MLPerf v1.1~\cite{mlperf} BERT on an AMD Instinct\texttrademark\ MI210 accelerator (GPU)~\cite{MI210} and scale its operation times based on changing hyperparameters and setup (e.g., sliced GEMM). This is beneficial as it helps us evaluate larger futuristic models (Transformer models are similar differing only in layers size/counts~\cite{PatiAga2021-demystifying,moolchandani2023amped}) and takes into account several GPU optimizations for Transformers~\cite{dao2022flashattention, mlperf_2_0_nvidia} already in MLPerf implementations.
Our projections further match those measured by prior works.
For example, AR's percentage runtime contribution projected for Mega-GPT-2 with TP-16 matches prior works' measurements on a similar system configuration~\cite{klenk2020network}.

\subsection{Applications, Deployment \& GEMMs}
\label{sec:eval-app-setup}

\noindent
\textbf{Models and their deployment}:
Since Transformers are fast-evolving, we evaluate \DESIGN{}'s impact on a range of Transformer models and TP degrees (Table~\ref{tab:models}).
For Megatron-GPT-2 (Mega-GPT-2)~\cite{ShoeybiPatwary2019-megatronlm} and T-NLG~\cite{Microsoft2020-tnlg} we use 16K and 8K input tokens (= input-length * batch-size) and TP degrees of eight and 16, given their modern intra-node setups~\cite{ShoeybiPatwary2019-megatronlm,Microsoft2020-tnlg,klenk2020network,jangda2022breaking}.
For larger Transformers like PALM~\cite{chowdhery2022palm}, GPT-3~\cite{BrownMann2020-gpt3}, and MT-NLG~\cite{smith2022using}) we use a higher slicing degree of 32 given their increasingly large memory capacity requirements~\cite{pati2023computation} and availability of nodes with larger device counts that can enable this slicing~\cite{GH200, JouppiYoon2021-tpuv4, tpu_pod}. We evaluate mixed-precision training which entails half-precision (FP16) forward and backpropagation and single-precision (FP32) weight updates. Similarly, we evaluate FP16 inference.

\noindent
\textbf{GEMMs}: GEMMs from the aforementioned applications are simulated using implementations from state-of-the-art BLAS libraries~\cite{cutlass_nvidia,cublas}.
Most GEMMs (including all GEMMs we evaluate) use a tiled GEMM implementation where each WG generates a complete tile of data (other implementations discussed in Section~\ref{subsec:disc-other-impl}).
Further, we evaluate GEMMs with both non-transposed (e.g., backward GEMMs) and transposed (e.g., forward GEMMs) input tensors, as observed in MLPerf's BERT~\cite{MattsonCheng2019-mlperfTrain, ReddiCheng2020-mlperfInfer}.

\subsection{Configurations}
\label{sec:eval-app-configs}

To evaluate \DESIGN{}'s efficacy we use the following configurations:

\begin{itemize}
  \item \textbf{\textit{Sequential}}: is the baseline configuration. Like modern systems, sequential executes sliced GEMMs and the following AR kernels sequentially.

  \item \textbf{\textit{\DESIGN}}: is our proposal which fuses and overlaps GEMM with RS (as described in Section~\ref{sec:fired_proposal}), followed by sequential all-gather (AG).

  \item \textbf{\textit{\DESIGN{}-MCA}}: uses fused GEMM-RS as in \DESIGN{}, but also includes the memory controller arbitration (MCA) discussed in Section~\ref{subsec:t3_mca}.

  \item \textbf{\textit{Ideal-GEMM-RS-Overlap}}: represents ideal GEMM and RS overlap in software.  Thus, its performance is the maximum of the GEMM's and the RS's isolated kernel execution times, followed by the AG time.  Moreover, it assumes no dependency constraints or resource contention between GEMM and RS.

  \item \textbf{\textit{Ideal-RS+NMC}}: uses RS with near-memory computing, which can provide additional speedup beyond a perfect overlap.  Thus, its performance is max(GEMM, RS+NMC) over Ideal-GEMM-RS-Overlap.

\end{itemize}

%% file: tables/sim-setup.tex
\begin{tabular}{|cl|}
\hline
\multicolumn{2}{|c|}{\textbf{System}}                  \\ \hline
\multicolumn{1}{|c|}{\#GPUs} & 8, 16                   \\ \hline
\multicolumn{1}{|c|}{\begin{tabular}[c]{@{}l@{}}Inter-GPU\\ Interconnect\end{tabular}} & \begin{tabular}[c]{@{}l@{}}Ring, 150 GB/s bi-directional  \\ 500 ns link latency \end{tabular}              \\ \hline
\multicolumn{2}{|c|}{\textbf{Per-GPU Config}}          \\ \hline
\multicolumn{1}{|c|}{\#CUs}  & 80, 1.4 GHz             \\ \hline
\multicolumn{1}{|c|}{Per-CU Config}          & \begin{tabular}[c]{@{}l@{}}2K threads, 128KB unified LDS  + L1 cache \\(with no write-allocate), 256KB RF  \end{tabular} \\ \hline
\multicolumn{1}{|c|}{L2}     & 16MB, 64 banks, 1.4 GHz \\ \hline
\multicolumn{1}{|c|}{HBM2}    & \begin{tabular}[c]{@{}l@{}}1 TB/s, 1 GHz, CCDWL=4,\\ Bank Grp.=4, rest ~\cite{chatterjee2017architecting} \end{tabular} \\ \hline
\end{tabular}

%% file: 07-evaluation.tex
\section{Results}
\label{sec:res}

\subsection{Execution Time Distribution \& Speedups}
\label{sec:t3_layer_speedups}

Figures~\ref{fig:t3_layer_distr} and~\ref{fig:t3_layer_speedup} show results for all sliced sub-layers in Transformers which require an AR: output projection (OP) and fully-connected-2 (FC-2) in forward pass (fwd) and fully-connected-1 (FC-1) and input projection (IP) in backprop (bwd).
We show these for Mega-GPT-2 and T-NLG, as well as two TP setups (TP of 8 and 16). Figure~\ref{fig:t3_layer_distr} shows each case's runtime distribution between the GEMM, RS, and AG. Figure~\ref{fig:t3_layer_speedup} shows their speedup over \textit{sequential} using \textit{\DESIGN{}}, \textit{\DESIGN{}-MCA}, as well as their speedups assuming an ideal overlap of GEMM with RS (\textit{Ideal-GEMM-RS-Overlap}) and additional speedups resulting from a faster RS with NMC (\textit{Ideal RS+NMC}).

\noindent
\subsubsection{Ideal Speedups} 
\begin{figure}[t!]
    \centering
    \includegraphics[width=\columnwidth,trim={1.5cm 11cm 25cm 0cm}]{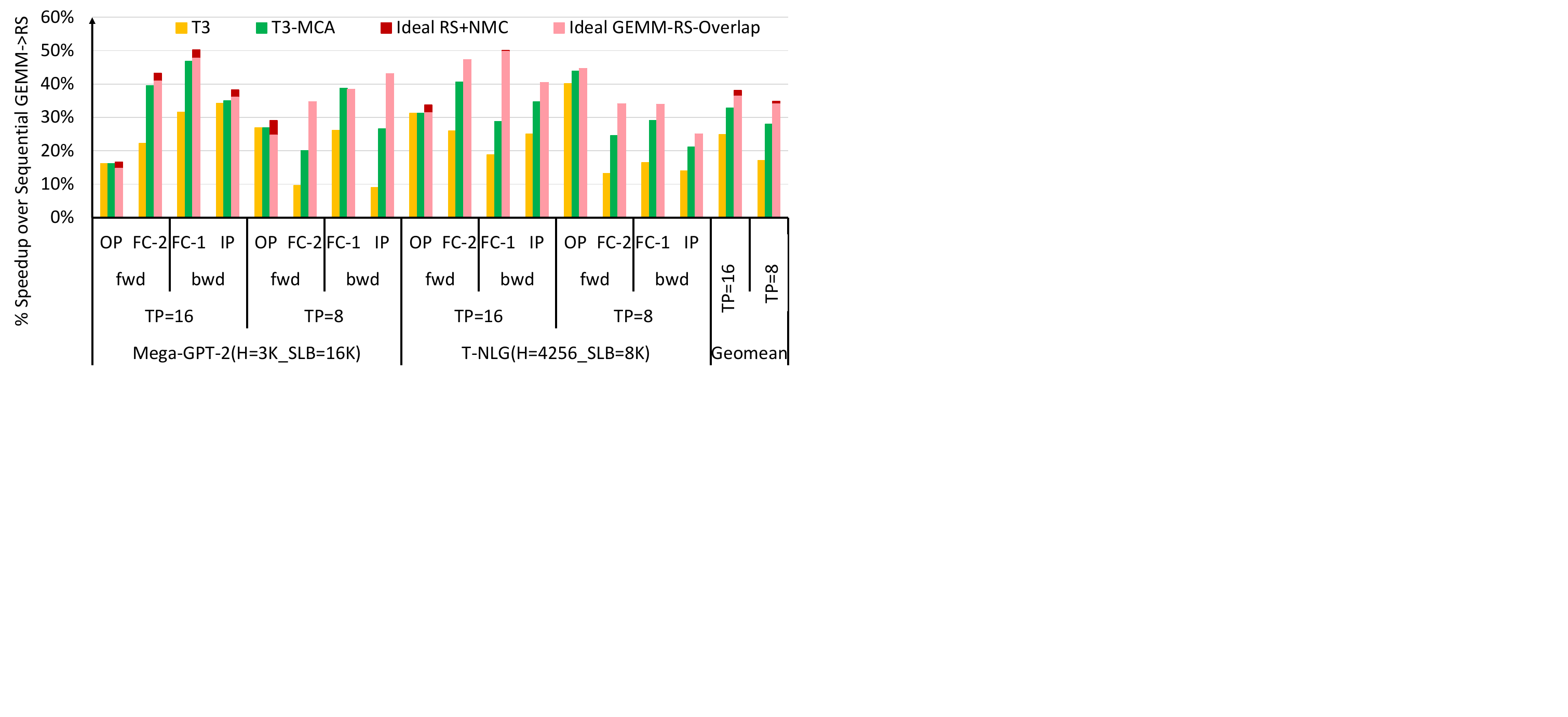}
   \vspace{-5ex}
    \caption{Transformer sub-layer speedups with \DESIGN{}}.
    \vspace{-3ex}
    \label{fig:t3_layer_speedup}
\end{figure}

Figure~\ref{fig:t3_layer_speedup} shows the ideal possible speedups and breaks them into two parts: first from overlapping the GEMM and RS kernels (\textit{Ideal-GEMM-RS-Overlap}) and second from improved RS performance due to NMC (\textit{Ideal RS+NMC}).

\begin{figure}[t!]
    \centering
    \includegraphics[width=\columnwidth, trim={0.7cm 11cm 14.5cm 4.5cm}]{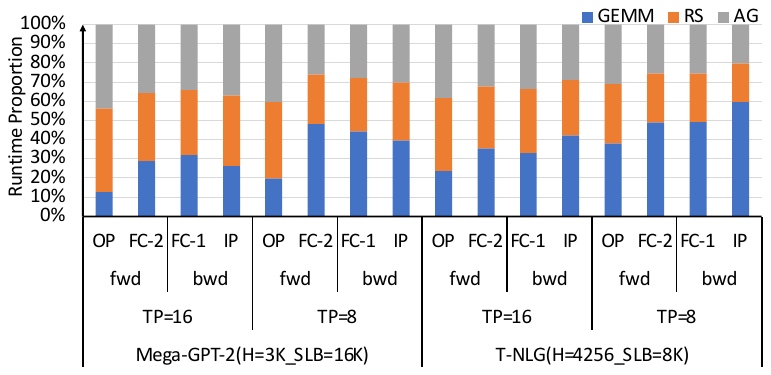}
    \caption{Transformer sub-layer runtime distribution.}
    \vspace{-1ex}
    \label{fig:t3_layer_distr}
\end{figure}

In Figure~\ref{fig:t3_layer_speedup} Ideal-GEMM-RS-Overlap (without resource and data-dependency constraints) shows considerable benefits from overlapping the producer GEMM and following RS: 50\% max speedup and 35\% geomean versus Sequential. %over their sequential execution.
Speedups vary both within and across models and depend on the isolated execution times of GEMM and RS (Figure~\ref{fig:t3_layer_distr}).
The situations where the GEMM and RS runtimes are similar (similar proportions in Figure~\ref{fig:t3_layer_distr}) have the maximum potential since the GEMM hides all of RS's cost.
For example, FC-1 in T-NLG with TP=16 obtains 50\% speedup.
Alternatively, the cases in which the GEMM and RS times are skewed show the least benefit since most of the GEMM or RS cost is exposed.
For example, Ideal-GEMM-RS-Overlap speedup is only 15\% for OP in Mega-GPT with TP=16.
However, the latter is uncommon and is a consequence of slicing a very small layer (OP is the smallest among all).
It does not hold for other sub-layers within the same model, or larger models as shown in the figures (also see Section~\ref{sec:eval_model_scale}). 
For a given hardware setup, these execution time ratios, and thus Ideal-GEMM-RS-Overlap speedups are dictated by layer parameters~\cite{pati2023computation}.

\begin{table}[tb!]
  \centering
  \resizebox{0.9\columnwidth}{!}{%
  \input{tables/models}
  }
  \caption{Studied models, their hyperparameters \& setup.}
  \vspace{-5ex}
  \label{tab:models}
\end{table}

In Figure~\ref{fig:t3_layer_speedup} Ideal-RS+NMC shows that additional speedup is possible beyond what perfect overlap provides.
Besides freeing all the CUs for GEMMs, performing RS reductions near memory also lowers RS's memory traffic (described in Section~\ref{subsubsec:t3_pim}).
This speeds up RS by 7\% and 3\% with TP=8 and TP=16, respectively.
NMC only reduces RS's final step time as interconnect costs dominate all prior steps and thus its runtime benefit decreases as TP, and thus total steps, increases.
As shown in Figure~\ref{fig:t3_layer_speedup}, this faster RS can reduce overlapped time and provide additional speedups of up to 4\%. Intuitively, the impact of a faster RS is only evident in layers in which RS is longer running than GEMM and is otherwise hidden when overlapped.

\noindent
\subsubsection{\DESIGN{} Speedups}
\label{sec:t3_speedups}

\DESIGN{} transparently overlaps GEMMs with their corresponding consumer RS in a fine-grained manner.
Moreover, \DESIGN{}'s lightweight track-\&-trigger mechanism and use of near-memory compute frees all CUs for GEMMs and reduces DRAM traffic (Figure~\ref{fig:t3_layer_dram} and Section~\ref{sec:eval_dram}), respectively.
Thus, \DESIGN{} achieves speedups of up to 39\% (20\% geomean, yellow bars,  Figure~\ref{fig:t3_layer_speedup}). 

Individual speedups vary considerably and are largely impacted by the extent of contention between DRAM traffic from the GEMM and the concurrent, RS  (details in Section~\ref{sec:eval_dram}). 
For OP layers, \DESIGN{} 
achieves close to the Ideal-GEMM-RS-Overlap speedups, and even exceeds them in certain cases.
This happens because the OP GEMMs are small and fit largely in the LLC, having very small DRAM read traffic in Sequential (shown in Figure~\ref{fig:t3_layer_dram}).
Thus, the additional DRAM traffic from the overlapped RS in \DESIGN{} has little impact on the GEMMs' progress/execution.
Instead, \DESIGN{} further improves RS runtimes in these cases via NMC and enables part of the additional Ideal-RS+NMC speedups.
Finally, although the track \& trigger mechanism operates at a small WF granularity, generally data from multiple WFs/WGs of a GEMM stage are ready to be sent concurrently, resulting in high network bandwidth utilization.
Furthermore, even when this is not true, \DESIGN{} can tolerate this because compute/GEMM execution and communication are overlapped, hiding the latency.

\begin{figure}[t!]
    \centering
    \includegraphics[width=0.7\columnwidth, trim={3cm 4cm 13.5cm 3cm}]{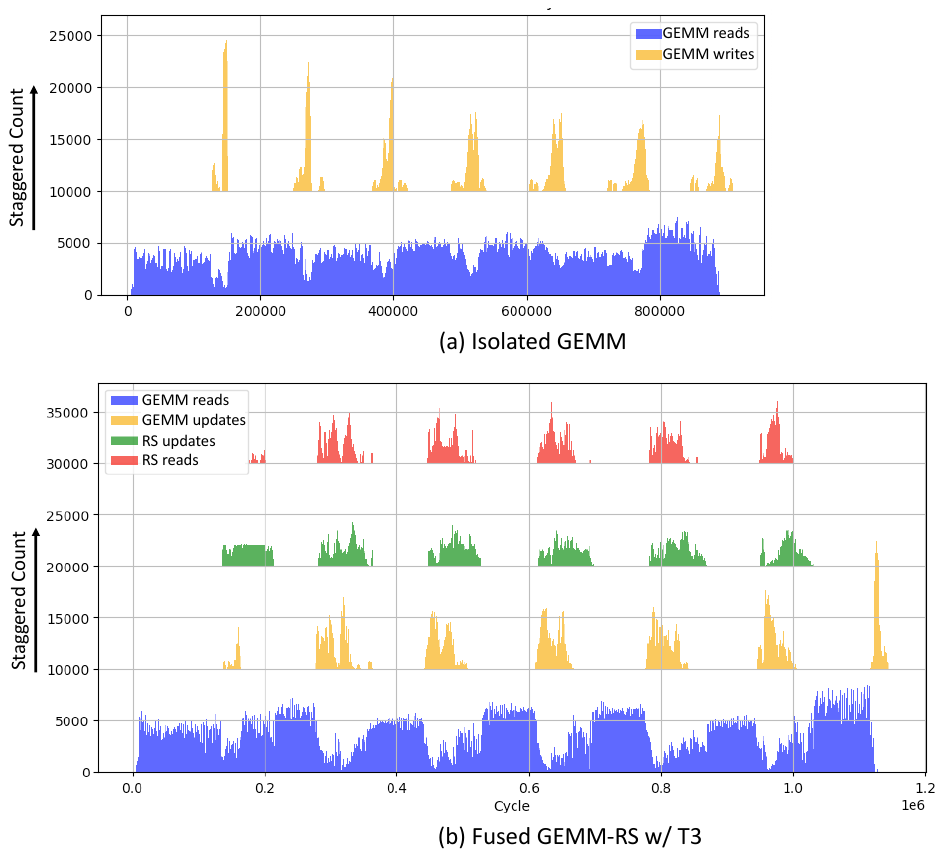}
    \caption{Overall DRAM traffic in (a) baseline GEMM, (b) \DESIGN{}, for T-NLG FC-2 with TP=8 and SLB=4K.}
    \label{fig:t3_dram_contention}
    \vspace{-2ex}
\end{figure}

In many other cases, and especially the much larger FC layers, the benefits are far from those with Ideal-GEMM-RS-Overlap (>15\% slower).
Figure~\ref{fig:t3_dram_contention} shows the DRAM traffic (Y-axis) and the GEMM slowdown (X-axis) with fine-grained overlapping, compared to the GEMM's isolated execution.
An isolated GEMM as shown in Figure~\ref{fig:t3_dram_contention}(a) executes in multiple stages (Section~\ref{sec:motiv_stages}), each with a read phase (blue) followed by a bursty write phase, which limit read traffic.
Overlapping RS induces additional DRAM traffic, as shown in Figure~\ref{fig:t3_dram_contention}(b).
Besides additional traffic, in \DESIGN{}, GEMM, and RS writes directly update memory using NMC (Section~\ref{subsubsec:t3_pim}).
These additional bursts of reads (RS\_reads for a stage are issued as soon as both the local and neighbors' copies have updated the memory) and updates (RS\_updates for the next stage from the previous neighbor) can further stall local GEMM reads as shown, causing GEMM to slow down considerably. 

\noindent
\subsubsection{\DESIGN{}-MCA Speedups} 
\label{sec:t3_mca_speedups}

\DESIGN{}-MCA (Section~\ref{subsec:t3_mca}) limits GEMM reads stalls due to bursty RS traffic (Section~\ref{sec:t3_speedups}, Figure~\ref{fig:t3_dram_contention}) using a simple arbitration logic. It prevents RS traffic from completely occupying DRAM queues by limiting communication-related accesses when a DRAM queue occupancy reaches a threshold (5, 10, 30, or no limit) determined by the memory intensity of the GEMM kernel.
\DESIGN{}-MCA provides considerable benefits over sequential execution; maximum of 47\%  and geomean of 30\% (29\% maximum and 13\% geomean over \DESIGN{}).
Furthermore, the geomean speedup with \DESIGN{}-MCA is only 5\% smaller than Ideal-GEMM-RS-Overlap.
There are individual cases where \DESIGN{}-MCA is far from ideal (e.g., FC-1 in T-NLG with TP=16).
These represent cases where L2 bypassing (for near-memory update) of GEMM writes hurts the GEMM's performance.
Consequently, the overall overlapped runtime also increases.

\subsection{Data Movement Reductions}
\label{sec:eval_dram}

Besides improved performance, \DESIGN{} and \DESIGN{}-MCA also reduce data movement to and from DRAM by a maximum of 36\% and an average of 22\% for the sub-layers.
Figure~\ref{fig:t3_layer_dram} shows the total memory accesses and their detailed breakdown (amongst GEMM, RS and AG reads/writes) for a single GPU across all cases.
While the AG reads/write remain constant between baseline (sequential) and \DESIGN{}-MCA, there is a combination of reasons which impact the rest: (a) fusion of GEMM and RS eliminates local writes from GEMM's first stage and reads from RS's first step, (b) near-memory reductions eliminate reading of partial copies every RS step, as well as the reads and writes in the final step's reduction, and (c) LLC bypassing of GEMM's output writes improves input read caching for cache-sensitive GEMMs, reducing GEMM's local reads. These impacts also vary depending on the TP degree: the one-time reductions (in the first and last RS step) have a much higher impact with smaller TP degrees due to fewer overall RS steps. Conversely, GEMM read caching impact is higher with a larger TP degree; larger TP/slicing leads to smaller, more LLC-amenable GEMMs. Overall, RS's reads reduce by 2.4$\times$ geomean (2.5$\times$ for TP=8, 2.2$\times$ for TP=16), both GEMM's and RS's writes reduce by 10\% geomean (14\% for TP=8, 7\% for TP=16), and finally GEMM's reads decrease by 1.56$\times$ geomean (1.2$\times$ for TP=8, 2$\times$ for TP=16).

\subsection{End-to-end Model Speedups}
As shown in Figure~\ref{fig:t3_future_models_e2e}, \DESIGN{} and \DESIGN{}-MCA speed up model training by a maximum of 9\% and 12\%, and geomean of 7\% and 10\%, respectively. Benefits are higher at larger TPs due to the overall higher proportion of the sliced sub-layers requiring AR (Section~\ref{fig:motiv-ar-percent}). Similarly, prompt processing and/or large input token processing during inference is also sped up by a maximum of 12\% and 15\%, and geomean of 9\% and 12\% with \DESIGN{} and \DESIGN{}-MCA, respectively. Inference speedups are better due to the overall higher proportion of sliced sub-layers resulting from no backprop compute. Finally, the MLPerfv1.1 implementation we evaluate does not include a key fusion optimization~\cite{dao2022flashattention}, which makes the non-sliced attention operations a significant 40-45\% of execution time.
Thus, we expect \DESIGN{}'s and \DESIGN{}-MCA's benefits to be much higher for newer MLPerf implementations. 

\subsection{Impact on Larger Transformers}
\label{sec:eval_model_scale}

We also evaluate larger Transformers with higher TP degrees as shown in Figure~\ref{fig:t3_future_models_e2e}.
Similar to the smaller models, layer-level speedups are high; max 35\% and geomean of 29\% for GPT-3 (175B parameters), PALM (530B  parameters), and MT-NLG (540B  parameters). These lead to up to 12\% and 14\% end-to-end speedup in their training and prompt phase of inference, respectively.
Thus, \DESIGN{}-MCA also effectively speeds up larger models.

%% file: tables/models.tex
\begin{tabular}{|c|c|c|c|}
\hline
\textbf{Model Name} & \textbf{Hyperparams} & \textbf{Inputs} & \textbf{TP degree} \\ \hline
Mega-GPT-2          & H=3072, L=74           & SL=1K, B=16     & 8, 16              \\ \hline
T-NLG               & H=4256, L=78           & SL=1K, B=8      & 8, 16              \\ \hline
GPT-3               & H=12K, L=96            & SL=1K, B=2      & 32                 \\ \hline
PALM                & H=18K, L=118            & SL=1K, B=2      & 32                 \\ \hline
MT-NLG              & H=20K, L=105            & SL=1K, B=2      & 32                 \\ \hline
\end{tabular}%

%% file: 08-discussion.tex
\section{Discussion}

\subsection{Other Collectives Implementation \& Types}
\label{subsec:disc-other-collectives}

\DESIGN{} supports other collectives and implementations via the configuration of GEMM's output address space (Section~\ref{subsec:t3_addr_config}).

\noindent
\textbf{Other implementations}: Collectives can have multiple implementations optimized for different topologies.
We focus on ring since it is commonly used in intra-node setups where tensor slicing is employed~\cite{nccl-tree-ring}.
\DESIGN{} can also support the \textit{direct} RS implementation in a fully-connected topology.
At every GEMM stage, the output from each device is scattered across the remaining devices using dedicated links and reduced at the destination. This is accomplished by changing the configuration in Figure~\ref{fig:t3_ring_addr_map_config} to slice each GEMM stage output and remote\_map each slice to a remote device. In this case \DESIGN{} eliminates memory accesses by the collective as it is completely orchestrated using GEMM stores. Similarly, it can also support other, inter-node implementations via appropriate programming of the track \& trigger mechanism.

\begin{figure}[t!]
    \centering
    \includegraphics[width=\columnwidth, trim={2cm 10cm 11.25cm 4.5cm}]{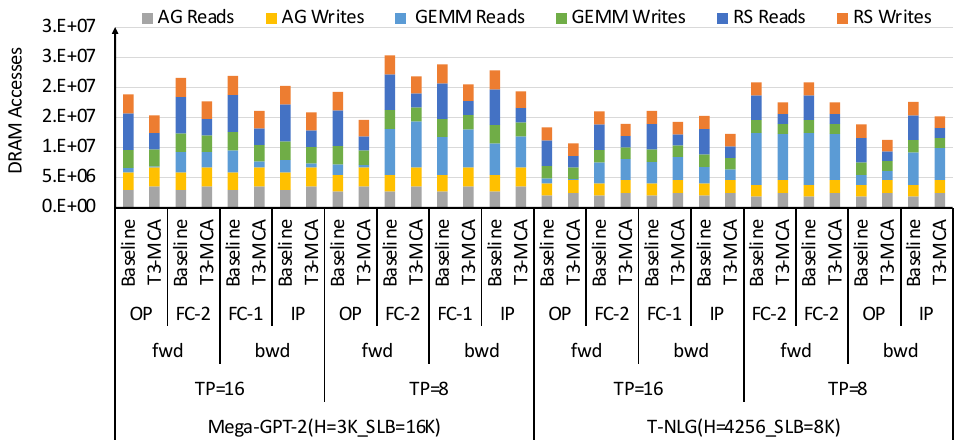}
    \caption{DRAM access per sub-layer.}
    \vspace{-3ex}
    \label{fig:t3_layer_dram}
\end{figure}

\noindent
\textbf{Other types}: Similarly, \DESIGN{} also supports other collectives.
A ring/direct all-gather (AG) reuses ring-RS's configuration and setup, except the GEMMs and DMA transfers do not update memory locations.
Similar to AG, \DESIGN{} can also support an all-to-all collective where devices exchange sub-arrays, except here the remote/dma\_mapped GEMM output is not written to local memory.

\begin{figure*}[!htb]
\minipage{0.6\textwidth}
    \includegraphics[width=1\columnwidth, trim={1cm 11cm 11.25cm 5cm}]{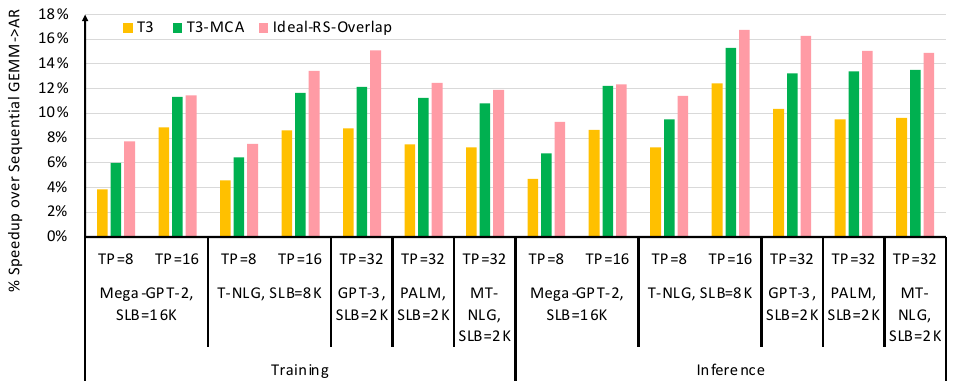}
    \caption{End-to-end model speedups.}
    \label{fig:t3_future_models_e2e}
\endminipage\hfill
\minipage{0.39\textwidth}%
    \includegraphics[width=\columnwidth, trim={0.5cm 13cm 18.5cm 5cm}]{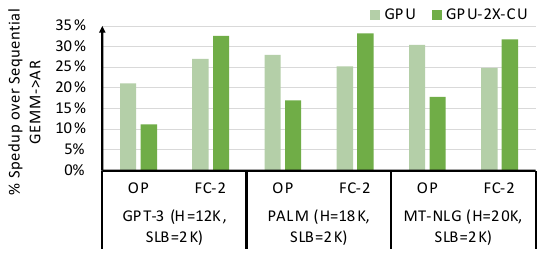}
    \caption{\DESIGN{} on future hardware with 2$\times$ compute.}
    \label{fig:t3_future_hardware_revised}
    \vspace{1ex}
\endminipage
\end{figure*}

\subsection{Other Distributed Techniques}
\label{subsec:disc-distr-techniques}

Although we focus on communication in tensor-parallel (TP) setups, \DESIGN{} is also applicable in other distributed setups where a producer's output is communicated via a collective.

\noindent
\textbf{Expert Parallelism}: Similar to TP, expert parallelism in Mixture-of-experts (MoEs)~\cite{rajbhandari2022deepspeed,Fedus21}
require serialized all-to-all communication which can be fused with \DESIGN{} as discussed in Section~\ref{subsec:disc-other-collectives}.

\noindent
\textbf{Data \& Pipeline Parallelism}:
\DESIGN{} also applies to data-parallel and pipeline-parallel setups which require RS, and peer-to-peer transfers, respectively.
While \DESIGN{}'s overlapping benefits may not provide additional benefits in such cases (these communications can be overlapped with other independent kernels), \DESIGN{}'s NMC and MCA techniques can help reduce memory bandwidth contention in these cases as well.

\noindent
\textbf{TP with All-gather}:
\DESIGN{} can be extended for distributed setups where the collective's output requires overlapping with a long-running consumer operation.
This is required if the producer is short-running (e.g., TP which all-gather's activations). Overlapping collective-consumer pairs is similar in principle to overlapping producer-collective and requires similar tracking/triggering mechanisms.
The Tracker would track “all-gathered-input$\rightarrow$GEMM-WG” instead of “GEMM-WG$\rightarrow$all-reduced-output”.
Moreover, instead of triggering a DMA, it would trigger a WG scheduling event (such as in Lustig \& Martonosi~\cite{lustig-full-empty}).
This can be challenging since the “all-gathered-input$\rightarrow$GEMM-WG” mapping can be kernel implementation dependent. 
However, additional programming hints could overcome this.

\subsection{Generative Inference}
While we focus on the communication-heavy training and prompt phase of inference, \DESIGN{} is also applicable in the generation phase of inference. Due to smaller input token counts (Section~\ref{subsec:bkg-transformers}), these phases are bound by memory accesses of model weights and can benefit from the aggregate memory bandwidth of multiple devices that TP provides~\cite{aminabadi2022deepspeed}. The resulting all-reduce of activations, while smaller than those in training and thus potentially latency-bound (due to small token counts), can still be overlapped and hidden with GEMM executions using \DESIGN{}.

\subsection{Other Reduction Substrates}
\label{subsec:disc-other-substrate}
While \DESIGN{} leverages NMC for atomic updates required in reduction-based collectives (e.g., RS, AR), it is not a requirement.
Such updates could also be handled via system-wide atomics on uncached data
without significant loss in performance.
Similarly, \DESIGN{} can also leverage switches for reductions as shown by prior works~\cite{klenk2020network}. 

\begin{table*}[tb!]
  \centering
  \input{tables/related}
  \caption{Comparing \DESIGN{}-MCA to prior work.}
  \label{tab:related}
  \vspace{-3ex}
\end{table*}

\subsection{Future Hardware \& Lower Precision}
\label{sec:eval_hardware_scale}

Since compute FLOPS have scaled much more than network link bandwidths across hardware generations~\cite{comp_bw_scaling,v100,a100,h100}, communication will likely be a larger proportion of the end-to-end execution both in more recent systems than the one we evaluate and in the future. Similarly, lowering precision~\cite{MSFT-BFP20,IntelFP822} decreases compute time much more (quadratically) than communication (linearly).
Thus, the benefits of hiding communication with techniques like \DESIGN{} will also apply to other GPU configurations and datatypes besides 16b.

To evaluate \DESIGN{}'s hiding capability in future systems, we study a system configuration where compute FLOPS scale more than network link bandwidth (2$\times$), which we term GPU-2X-CU. While the scaling GPU FLOPs across generations largely result from more powerful CUs (larger/faster tensor processing), we simulate it by scaling the number of CUs and keeping the underlying network the same. This enables us to use the latest/validated GPU model and GEMM traces that Accel-Sim supports~\cite{KhairyShen2021-accelSim}. Figure~\ref{fig:t3_future_hardware_revised} shows that for larger layers (FC-2) where compute time dominates, compute becomes faster with 2$\times$ CUs which lowers the compute:communication ratio across the models. This shortens the critical path and leads to larger benefits with overlapping compute and communication with \DESIGN{}. Conversely, for smaller layers (OP), where compute and communication are more balanced, faster compute exposes communication on critical path, lowering \DESIGN{}'s benefits. Note that, for such scenarios, communication optimizations will be necessary~\cite{cai2021synthesizing,shah2023taccl}. Nevertheless, the larger layers have a more prominent impact on overall execution and for these, \DESIGN{}'s benefits only improve.

\subsection{NMC for Following Operations}

Collectives, specifically all-reduce in Transformers, are usually followed by other memory-intensive operations on all devices (e.g., parameter updates in DP~\cite{PatiAga2021-demystifying} or residual/dropout layers in TP).
These operations redundantly operate on the entire all-reduced array on each device. With \DESIGN{}, these following memory-intensive operations can be executed using NMC~\cite{PatiAga2021-demystifying} on (reduced) sub-arrays of data before they are all-gathered/broadcasted to the remaining devices, thus reducing redundancy, and further accelerating distributed Transformer models.

\subsection{Other GEMM Implementations}
\label{subsec:disc-other-impl}

\DESIGN{} focuses on the most common tiled GEMM implementation with a WG/WF responsible to generate an entire tile/sub-tile of data.
However, \DESIGN{} can support other implementations, such as split-K~\cite{splitk}.
A split-K implementation slices work in the accumulation or $K$ dimension, such that multiple WGs are responsible for a single tile, each generating a partial tile that is reduced after.
Split-K increases parallelism when the output size ($MxN$) is small but the $K$ dimension is large.
However, tensor-sliced GEMMs, which require AR, have large output sizes and small $K$ dimensions.
Naively, \DESIGN{} with a split-K implementation (with more than one update to an element) will cause multiple local and remote updates per memory location.
To prevent this, \DESIGN{} can use the kernel packets’ tile-size metadata to deduce split-k degree (=(\#WGs * tile-size)/(M*N)), i.e., the number of updates per element.
The virtual addresses in the tracker (Section~\ref{sec:t3_tracker}) can be used to determine WFs/WGs/tracker entries that update the same tile, allowing the tracker to trigger remote DMA only after all updates to the tile are complete. 

\subsection{Multi-node Setups}
\label{subsec:disc-other-multiNode}

Tensor-parallelism, with serialized communication is usually employed within a node, which generally has high-speed homogeneous links.
However, \DESIGN{} can also be applied to serialized communication in inter-node setups with slower and often heterogeneous links.
Consequently, communication costs can be much larger than GEMM executions, potentially limiting the benefits from fine-grained overlap: once the computation is completely overlapped, the remaining communication costs will be exposed~\cite{wang2022overlap}.
Nevertheless, \DESIGN{} can still provide benefits from hiding the GEMM execution cost as much as possible.

%% file: tables/related.tex
{\footnotesize
\begin{tabular}{|c|c|c|c|c|c|c|}
\hline
\textbf{\begin{tabular}[c]{@{}c@{}} Approach /\\ Features \end{tabular}}                   & \textbf{\begin{tabular}[c]{@{}c@{}}GPU\\Support\end{tabular}} & \textbf{\begin{tabular}[c]{@{}c@{}c@{}}Transparent\\Communication \end{tabular}} & \textbf{\begin{tabular}[c]{@{}c@{}}Overlap\\Communication\end{tabular}} & \textbf{\begin{tabular}[c]{@{}c@{}}Reduce\\Contention\end{tabular}} & \textbf{\begin{tabular}[c]{@{}c@{}}No Additional\\Accelerator\end{tabular}}  & \textbf{\begin{tabular}[c]{@{}c@{}}Topology-\\independent \end{tabular}}  \\ \hline

\textbf{In-switch~\cite{klenk2020network}}        & \textcolor[HTML]{00A99A}{\checkmark}                      & \textcolor{red}{X}                                                                                    & \textcolor{red}{X}                                                                                      &  \textcolor[HTML]{00A99A}{\checkmark}                                                                                 & \textcolor{red}{X}                                                                    & \textcolor{red}{X}   \\ \hline
\textbf{ACE~\cite{rashidi2021enabling}}        & \textcolor[HTML]{00A99A}{\checkmark}                      & \textcolor{red}{X}                                                                                    & \textcolor{red}{X}                                                                                     &  \textcolor[HTML]{00A99A}{\checkmark}                                                                                 & \textcolor{red}{X}     & \textcolor{red}{X}                                                                  \\ \hline
\textbf{CoCoNet~\cite{jangda2022breaking}}               & \textcolor[HTML]{00A99A}{\checkmark}                     &         \textcolor{red}{X}                               & \textcolor[HTML]{00A99A}{\checkmark}                                                                                          & \textcolor{red}{X}                                                                    & \textcolor[HTML]{00A99A}{\checkmark}       & \textcolor[HTML]{00A99A}{\checkmark}                                                             \\ \hline
\textbf{Google Decomposition~\cite{wang2022overlap}}     & \textcolor{red}{X}        & \textcolor{red}{X}                                                                                      & \textcolor[HTML]{00A99A}{\checkmark}                                                                                           & \textcolor{red}{X}                                                                           & \textcolor[HTML]{00A99A}{\checkmark}             & \textcolor[HTML]{00A99A}{\checkmark}                                                          \\ \hline
\textbf{\begin{tabular}[c]{@{}c@{}}\DESIGN{}-MCA\end{tabular}}           & \textcolor[HTML]{00A99A}{\checkmark}                   & \textcolor[HTML]{00A99A}{\checkmark}                                                                                   & \textcolor[HTML]{00A99A}{\checkmark}                                                                                          & \textcolor[HTML]{00A99A}{\checkmark}                                                                            & \textcolor[HTML]{00A99A}{\checkmark}         & \textcolor[HTML]{00A99A}{\checkmark}                                                             \\ \hline
\end{tabular}
}

%% file: 09-related.tex
\vspace{-1ex}
\section{Related Work}
\label{sec:fired_related}

Table~\ref{tab:related} compares \DESIGN{}-MCA with prior works across several key metrics. 
Some prior work has designed \textit{in-switch collectives} to speed up communication by up to 2$\times$~\cite{klenk2020network}.
However, this cannot eliminate serialized communication from the critical path.
Furthermore, they are topology-dependent, requiring switches.
Enabling fine-grained overlap of compute and communication is essential to effectively hide the cost of communication.
Existing attempts to do this, like \textit{CocoNet}\cite{jangda2022breaking} and \textit{Google Decomposition}~\cite{wang2022overlap}, have limitations.
\textit{Google Decomposition} requires changes to matrix multiplication (GEMMs) kernels which can be disruptive to GPU software infrastructure (Section~\ref{sec:chal_complex_sw}).

Furthermore, both approaches can suffer from hardware resource contention between compute and communication (Section~\ref{sec:chal-resource}).
Works that reduce contention only address coarse-grained overlap of compute and communication in cases like DP, lacking support for fine-grained overlap in serialized collectives~\cite{rashidi2021enabling}.
Moreover, they rely on dedicated accelerators.
Other recent work fuses communication within the computation kernel to enable fine-grained overlap, such that a GPU kernel performs both computation and dependent communication at the WG level~\cite{punniyamurthy2023gpu}. 
However, this requires explicit changes to the compute kernels and is not readily applicable for collectives involving simple arithmetic operation such as reduce-scatter -- which will still be limited by inter-GPU synchronization. Finally, other work like Syndicate increases coarse-grained overlap opportunities and efficiency in distributed training.
However, Syndicate cannot hide serialized communication~\cite{mahajan2023better}. 
\textit{\DESIGN{}-MCA overcomes these shortcomings and achieves a transparent overlap of serialized communication with compute, while minimizing resource contention}.

%% file: 10-conclusion.tex
\section{Conclusion}
\label{sec:conc}

Transformer models increasingly rely on distributed techniques, requiring communication between multiple devices.
This communication can limit scaling efficiency, especially for techniques like Tensor Parallelism (TP) which serialize communication with model execution. While a fine-grained overlap of the serialized communication with its producer computation can help hide the cost, realizing it with GPUs is challenging due to software complexities and resource contention between compute and communication. To overcome this, we propose \DESIGN{}, which transparently and efficiently fuses and overlaps serialized inter-device communication with the producer's compute. It orchestrates communication on the producer's stores by configuring the producer's output address space mapping and using a programmable track and trigger mechanism in hardware. This reduces application impact and also eliminates contention for GPU compute resources. \DESIGN{} additionally uses near-memory computing and a memory-controller arbitration policy to reduce memory-bandwidth contention. 
Overall, \DESIGN{} improves performance by 30\% geomean (max 47\%) and reduces data movement by 22\% geomean (max 36\%) over state-of-the-art approaches.
Moreover, \DESIGN{}'s benefits hold as models and hardware scale. 

%% file: 11-ack.tex
\section*{Acknowledgements}
We would like to thank our shepherd, Saeed Maleki, and the anonymous reviewers for their feedback that helped improve this paper.
We would also like to thank Mahmoud Khairy Abdallah for his help with the Accel-Sim simulator. This work is supported in part at the University of Wisconsin-Madison by the Vilas Life Cycle Professorship program and a Fall Research Competition grant, as well as the National Science Foundation under grant ENS-1925485.
AMD, AMD Ryzen, AMD Radeon, and combinations thereof are trademarks of Advanced Micro Devices, Inc.
Other product names used in this publication are for identification purposes only and may be trademarks of their respective companies.

%% file: 00-sample-sigplan.bbl

\begin{thebibliography}{87}

%%% the Web address.

\ifx \showCODEN    \undefined \def \showCODEN     #1{\unskip}     \fi
\ifx \showDOI      \undefined \def \showDOI       #1{#1}\fi
\ifx \showISBNx    \undefined \def \showISBNx     #1{\unskip}     \fi
\ifx \showISBNxiii \undefined \def \showISBNxiii  #1{\unskip}     \fi
\ifx \showISSN     \undefined \def \showISSN      #1{\unskip}     \fi
\ifx \showLCCN     \undefined \def \showLCCN      #1{\unskip}     \fi
\ifx \shownote     \undefined \def \shownote      #1{#1}          \fi
\ifx \showarticletitle \undefined \def \showarticletitle #1{#1}   \fi
\ifx \showURL      \undefined \def \showURL       {\relax}        \fi
% invisible to TeX
\providecommand\bibfield[2]{#2}
\providecommand\bibinfo[2]{#2}
\providecommand\natexlab[1]{#1}
\providecommand\showeprint[2][]{arXiv:#2}

\bibitem[{AMD}(2018)]%
        {rccl}
\bibfield{author}{\bibinfo{person}{{AMD}}.} \bibinfo{year}{2018}\natexlab{}.
\newblock \bibinfo{title}{{AMD's ROCm Communication Collectives Library}}.
\newblock \bibinfo{howpublished}{"\url{https://github.com/ROCmSoftwarePlatform/rccl/wiki}"}.
\newblock


\bibitem[AMD(2019)]%
        {rocblas}
\bibfield{author}{\bibinfo{person}{AMD}.} \bibinfo{year}{2019}\natexlab{}.
\newblock \bibinfo{title}{{AMD's BLAS Library}}.
\newblock \bibinfo{howpublished}{"\url{https://github.com/ROCmSoftwarePlatform/rocBLAS}"}.
\newblock


\bibitem[{AMD}(2020)]%
        {Tensile}
\bibfield{author}{\bibinfo{person}{{AMD}}.} \bibinfo{year}{2020}\natexlab{}.
\newblock \bibinfo{title}{{AMD's tool for creating a benchmark-driven backend library for GEMMs}}.
\newblock \bibinfo{howpublished}{"\url{https://github.com/ROCmSoftwarePlatform/Tensile/}"}.
\newblock


\bibitem[{AMD}(2021)]%
        {amd-kernel-object}
\bibfield{author}{\bibinfo{person}{{AMD}}.} \bibinfo{year}{2021}\natexlab{}.
\newblock \bibinfo{title}{{AMD HSA Code Object Format}}.
\newblock \bibinfo{howpublished}{"\url{https://rocmdocs.amd.com/en/latest/ROCm_Compiler_SDK/ROCm-Codeobj-format.html}"}.
\newblock


\bibitem[{AMD}(2022)]%
        {MI210}
\bibfield{author}{\bibinfo{person}{{AMD}}.} \bibinfo{year}{2022}\natexlab{}.
\newblock \bibinfo{title}{{AMD INSTINCT\texttrademark\ MI210 ACCELERATOR}}.
\newblock \bibinfo{howpublished}{\url{https://www.amd.com/en/products/server-accelerators/amd-instinct-mi210}}.
\newblock


\bibitem[Aminabadi et~al\mbox{.}(2022)]%
        {aminabadi2022deepspeed}
\bibfield{author}{\bibinfo{person}{Reza~Yazdani Aminabadi}, \bibinfo{person}{Samyam Rajbhandari}, \bibinfo{person}{Ammar~Ahmad Awan}, \bibinfo{person}{Cheng Li}, \bibinfo{person}{Du Li}, \bibinfo{person}{Elton Zheng}, \bibinfo{person}{Olatunji Ruwase}, \bibinfo{person}{Shaden Smith}, \bibinfo{person}{Minjia Zhang}, \bibinfo{person}{Jeff Rasley}, {and} \bibinfo{person}{Yuxiong He}.} \bibinfo{year}{2022}\natexlab{}.
\newblock \showarticletitle{{DeepSpeed-Inference: Enabling Efficient Inference of Transformer Models at Unprecedented Scale}}. In \bibinfo{booktitle}{\emph{{Proceedings of the International Conference for High Performance Computing, Networking, Storage and Analysis}}} \emph{(\bibinfo{series}{SC})}. IEEE, \bibinfo{publisher}{IEEE Press}, \bibinfo{address}{Piscataway, NJ, USA}, \bibinfo{pages}{1--15}.
\newblock
\showISBNx{9784665454445}


\bibitem[Bao et~al\mbox{.}(2023)]%
        {BaoSun2023-naviSim}
\bibfield{author}{\bibinfo{person}{Yuhui Bao}, \bibinfo{person}{Yifan Sun}, \bibinfo{person}{Zlatan Feric}, \bibinfo{person}{Michael~Tian Shen}, \bibinfo{person}{Micah Weston}, \bibinfo{person}{Jos\'{e}~L. Abell\'{a}n}, \bibinfo{person}{Trinayan Baruah}, \bibinfo{person}{John Kim}, \bibinfo{person}{Ajay Joshi}, {and} \bibinfo{person}{David Kaeli}.} \bibinfo{year}{2023}\natexlab{}.
\newblock \showarticletitle{{NaviSim: A Highly Accurate GPU Simulator for AMD RDNA GPUs}}. In \bibinfo{booktitle}{\emph{{Proceedings of the International Conference on Parallel Architectures and Compilation Techniques}}} (Chicago, Illinois) \emph{(\bibinfo{series}{PACT '22})}. \bibinfo{publisher}{Association for Computing Machinery}, \bibinfo{address}{New York, NY, USA}, \bibinfo{pages}{333–345}.
\newblock
\showISBNx{9781450398688}
\urldef\tempurl%
\url{https://doi.org/10.1145/3559009.3569666}
\showDOI{\tempurl}


\bibitem[Benaich and Hogarth(2022)]%
        {StateofAI22}
\bibfield{author}{\bibinfo{person}{Nathan Benaich} {and} \bibinfo{person}{Ian Hogarth}.} \bibinfo{year}{2022}\natexlab{}.
\newblock \bibinfo{title}{{State of AI Report 2022}}.
\newblock \bibinfo{howpublished}{\url{https://www.stateof.ai/}}.
\newblock


\bibitem[Brown et~al\mbox{.}(2020)]%
        {BrownMann2020-gpt3}
\bibfield{author}{\bibinfo{person}{Tom Brown}, \bibinfo{person}{Benjamin Mann}, \bibinfo{person}{Nick Ryder}, \bibinfo{person}{Melanie Subbiah}, \bibinfo{person}{Jared~D Kaplan}, \bibinfo{person}{Prafulla Dhariwal}, \bibinfo{person}{Arvind Neelakantan}, \bibinfo{person}{Pranav Shyam}, \bibinfo{person}{Girish Sastry}, \bibinfo{person}{Amanda Askell}, \bibinfo{person}{Sandhini Agarwal}, \bibinfo{person}{Ariel Herbert-Voss}, \bibinfo{person}{Gretchen Krueger}, \bibinfo{person}{Tom Henighan}, \bibinfo{person}{Rewon Child}, \bibinfo{person}{Aditya Ramesh}, \bibinfo{person}{Daniel Ziegler}, \bibinfo{person}{Jeffrey Wu}, \bibinfo{person}{Clemens Winter}, \bibinfo{person}{Chris Hesse}, \bibinfo{person}{Mark Chen}, \bibinfo{person}{Eric Sigler}, \bibinfo{person}{Mateusz Litwin}, \bibinfo{person}{Scott Gray}, \bibinfo{person}{Benjamin Chess}, \bibinfo{person}{Jack Clark}, \bibinfo{person}{Christopher Berner}, \bibinfo{person}{Sam McCandlish}, \bibinfo{person}{Alec Radford}, \bibinfo{person}{Ilya Sutskever}, {and}
  \bibinfo{person}{Dario Amodei}.} \bibinfo{year}{2020}\natexlab{}.
\newblock \showarticletitle{{Language Models are Few-Shot Learners}}. In \bibinfo{booktitle}{\emph{{Advances in Neural Information Processing Systems}}} \emph{(\bibinfo{series}{NeurIPS}, Vol.~\bibinfo{volume}{33})}, \bibfield{editor}{\bibinfo{person}{H.~Larochelle}, \bibinfo{person}{M.~Ranzato}, \bibinfo{person}{R.~Hadsell}, \bibinfo{person}{M.~F. Balcan}, {and} \bibinfo{person}{H.~Lin}} (Eds.). \bibinfo{publisher}{Curran Associates Inc.}, \bibinfo{address}{Red Hook, NY, USA}, \bibinfo{pages}{1877--1901}.
\newblock
\showISBNx{9781713829546}


\bibitem[Cai et~al\mbox{.}(2021)]%
        {cai2021synthesizing}
\bibfield{author}{\bibinfo{person}{Zixian Cai}, \bibinfo{person}{Zhengyang Liu}, \bibinfo{person}{Saeed Maleki}, \bibinfo{person}{Madanlal Musuvathi}, \bibinfo{person}{Todd Mytkowicz}, \bibinfo{person}{Jacob Nelson}, {and} \bibinfo{person}{Olli Saarikivi}.} \bibinfo{year}{2021}\natexlab{}.
\newblock \showarticletitle{{Synthesizing Optimal Collective Algorithms}}. In \bibinfo{booktitle}{\emph{{Proceedings of the 26th ACM SIGPLAN Symposium on Principles and Practice of Parallel Programming}}} \emph{(\bibinfo{series}{PPOPP})}. \bibinfo{publisher}{Association for Computing Machinery}, \bibinfo{address}{New York, NY, USA}, \bibinfo{pages}{62--75}.
\newblock
\showISBNx{9781450382946}
\urldef\tempurl%
\url{https://doi.org/10.1145/3437801.3441620}
\showDOI{\tempurl}


\bibitem[Chatterjee et~al\mbox{.}(2017)]%
        {chatterjee2017architecting}
\bibfield{author}{\bibinfo{person}{Niladrish Chatterjee}, \bibinfo{person}{Mike O'Connor}, \bibinfo{person}{Donghyuk Lee}, \bibinfo{person}{Daniel~R Johnson}, \bibinfo{person}{Stephen~W Keckler}, \bibinfo{person}{Minsoo Rhu}, {and} \bibinfo{person}{William~J Dally}.} \bibinfo{year}{2017}\natexlab{}.
\newblock \showarticletitle{{Architecting an Energy-Efficient DRAM System for GPUs}}. In \bibinfo{booktitle}{\emph{{23rd IEEE International Symposium on High Performance Computer Architecture}}} \emph{(\bibinfo{series}{HPCA})}. IEEE, \bibinfo{publisher}{IEEE Computer Society}, \bibinfo{address}{Washington, DC, USA}, \bibinfo{pages}{73--84}.
\newblock


\bibitem[Chowdhery et~al\mbox{.}(2022)]%
        {chowdhery2022palm}
\bibfield{author}{\bibinfo{person}{Aakanksha Chowdhery}, \bibinfo{person}{Sharan Narang}, \bibinfo{person}{Jacob Devlin}, \bibinfo{person}{Maarten Bosma}, \bibinfo{person}{Gaurav Mishra}, \bibinfo{person}{Adam Roberts}, \bibinfo{person}{Paul Barham}, \bibinfo{person}{Hyung~Won Chung}, \bibinfo{person}{Charles Sutton}, \bibinfo{person}{Sebastian Gehrmann}, \bibinfo{person}{Parker Schuh}, \bibinfo{person}{Kensen Shi}, \bibinfo{person}{Sasha Tsvyashchenko}, \bibinfo{person}{Joshua Maynez}, \bibinfo{person}{Abhishek Rao}, \bibinfo{person}{Parker Barnes}, \bibinfo{person}{Yi Tay}, \bibinfo{person}{Noam Shazeer}, \bibinfo{person}{Vinodkumar Prabhakaran}, \bibinfo{person}{Emily Reif}, \bibinfo{person}{Nan Du}, \bibinfo{person}{Ben Hutchinson}, \bibinfo{person}{Reiner Pope}, \bibinfo{person}{James Bradbury}, \bibinfo{person}{Jacob Austin}, \bibinfo{person}{Michael Isard}, \bibinfo{person}{Guy Gur-Ari}, \bibinfo{person}{Pengcheng Yin}, \bibinfo{person}{Toju Duke}, \bibinfo{person}{Anselm Levskaya},
  \bibinfo{person}{Sanjay Ghemawat}, \bibinfo{person}{Sunipa Dev}, \bibinfo{person}{Henryk Michalewski}, \bibinfo{person}{Xavier Garcia}, \bibinfo{person}{Vedant Misra}, \bibinfo{person}{Kevin Robinson}, \bibinfo{person}{Liam Fedus}, \bibinfo{person}{Denny Zhou}, \bibinfo{person}{Daphne Ippolito}, \bibinfo{person}{David Luan}, \bibinfo{person}{Hyeontaek Lim}, \bibinfo{person}{Barret Zoph}, \bibinfo{person}{Alexander Spiridonov}, \bibinfo{person}{Ryan Sepassi}, \bibinfo{person}{David Dohan}, \bibinfo{person}{Shivani Agrawal}, \bibinfo{person}{Mark Omernick}, \bibinfo{person}{Andrew~M. Dai}, \bibinfo{person}{Thanumalayan~Sankaranarayana Pillai}, \bibinfo{person}{Marie Pellat}, \bibinfo{person}{Aitor Lewkowycz}, \bibinfo{person}{Erica Moreira}, \bibinfo{person}{Rewon Child}, \bibinfo{person}{Oleksandr Polozov}, \bibinfo{person}{Katherine Lee}, \bibinfo{person}{Zongwei Zhou}, \bibinfo{person}{Xuezhi Wang}, \bibinfo{person}{Brennan Saeta}, \bibinfo{person}{Mark Diaz}, \bibinfo{person}{Orhan Firat},
  \bibinfo{person}{Michele Catasta}, \bibinfo{person}{Jason Wei}, \bibinfo{person}{Kathy Meier-Hellstern}, \bibinfo{person}{Douglas Eck}, \bibinfo{person}{Jeff Dean}, \bibinfo{person}{Slav Petrov}, {and} \bibinfo{person}{Noah Fiedel}.} \bibinfo{year}{2022}\natexlab{}.
\newblock \showarticletitle{{PaLM: Scaling Language Modeling with Pathways}}.
\newblock \bibinfo{journal}{\emph{arXiv preprint arXiv:2204.02311}} (\bibinfo{year}{2022}), \bibinfo{numpages}{87}~pages.
\newblock


\bibitem[Dao et~al\mbox{.}(2022)]%
        {dao2022flashattention}
\bibfield{author}{\bibinfo{person}{Tri Dao}, \bibinfo{person}{Dan Fu}, \bibinfo{person}{Stefano Ermon}, \bibinfo{person}{Atri Rudra}, {and} \bibinfo{person}{Christopher R{\'e}}.} \bibinfo{year}{2022}\natexlab{}.
\newblock \showarticletitle{{FlashAttention: Fast and Memory-efficient Exact Attention with IO-Awareness}}.
\newblock \bibinfo{journal}{\emph{Advances in Neural Information Processing Systems}}  \bibinfo{volume}{35} (\bibinfo{year}{2022}), \bibinfo{pages}{16344--16359}.
\newblock


\bibitem[Devlin et~al\mbox{.}(2019)]%
        {DevlinChang18-bert}
\bibfield{author}{\bibinfo{person}{Jacob Devlin}, \bibinfo{person}{Ming{-}Wei Chang}, \bibinfo{person}{Kenton Lee}, {and} \bibinfo{person}{Kristina Toutanova}.} \bibinfo{year}{2019}\natexlab{}.
\newblock \showarticletitle{{BERT: Pre-training of Deep Bidirectional Transformers for Language Understanding}}. In \bibinfo{booktitle}{\emph{{Proceedings of the 2019 Conference of the North American Chapter of the Association for Computational Linguistics: Human Language Technologies}}} \emph{(\bibinfo{series}{NAACL-HLT})}, \bibfield{editor}{\bibinfo{person}{Jill Burstein}, \bibinfo{person}{Christy Doran}, {and} \bibinfo{person}{Thamar Solorio}} (Eds.). \bibinfo{publisher}{Association for Computational Linguistics}, \bibinfo{address}{Morristown, NJ, USA}, \bibinfo{pages}{4171--4186}.
\newblock
\urldef\tempurl%
\url{https://doi.org/10.18653/v1/n19-1423}
\showDOI{\tempurl}


\bibitem[Eassa and Eryilmaz(2022)]%
        {mlperf_2_0_nvidia}
\bibfield{author}{\bibinfo{person}{Shraf Eassa} {and} \bibinfo{person}{Sukru~Burc Eryilmaz}.} \bibinfo{year}{2022}\natexlab{}.
\newblock \bibinfo{title}{{The Full Stack Optimization Powering NVIDIA MLPerf Training v2.0 Performance}}.
\newblock \bibinfo{howpublished}{\url{https://developer.nvidia.com/blog/boosting-mlperf-training-performance-with-full-stack-optimization/}}.
\newblock


\bibitem[El~Hajj et~al\mbox{.}(2016)]%
        {ElHajjGomezLuna2016-klap}
\bibfield{author}{\bibinfo{person}{Izzat El~Hajj}, \bibinfo{person}{Juan Gomez-Luna}, \bibinfo{person}{Cheng Li}, \bibinfo{person}{Li-Wen Chang}, \bibinfo{person}{Dejan Milojicic}, {and} \bibinfo{person}{Wen-mei Hwu}.} \bibinfo{year}{2016}\natexlab{}.
\newblock \showarticletitle{{KLAP: Kernel Launch Aggregation and Promotion for Optimizing Dynamic Parallelism}}. In \bibinfo{booktitle}{\emph{{49th Annual IEEE/ACM International Symposium on Microarchitecture}}} \emph{(\bibinfo{series}{MICRO})}. IEEE, \bibinfo{publisher}{IEEE Press}, \bibinfo{address}{Piscataway, NJ, USA}, \bibinfo{pages}{1--12}.
\newblock
\urldef\tempurl%
\url{https://doi.org/10.1109/MICRO.2016.7783716}
\showDOI{\tempurl}


\bibitem[Fedus et~al\mbox{.}(2022)]%
        {Fedus21}
\bibfield{author}{\bibinfo{person}{William Fedus}, \bibinfo{person}{Barret Zoph}, {and} \bibinfo{person}{Noam Shazeer}.} \bibinfo{year}{2022}\natexlab{}.
\newblock \showarticletitle{{Switch Transformers: Scaling to Trillion Parameter Models with Simple and Efficient Sparsity}}.
\newblock \bibinfo{journal}{\emph{The Journal of Machine Learning Research}} \bibinfo{volume}{23}, \bibinfo{number}{1}, Article \bibinfo{articleno}{120} (\bibinfo{date}{jan} \bibinfo{year}{2022}), \bibinfo{numpages}{39}~pages.
\newblock
\showISSN{1532-4435}


\bibitem[Fousek et~al\mbox{.}(2011)]%
        {FousekFilipovivc2011-fuseGPUMap}
\bibfield{author}{\bibinfo{person}{Jan Fousek}, \bibinfo{person}{Ji\v{r}i Filipovi\v{c}}, {and} \bibinfo{person}{Matu\v{s} Madzin}.} \bibinfo{year}{2011}\natexlab{}.
\newblock \showarticletitle{{Automatic Fusions of CUDA-GPU Kernels for Parallel Map}}.
\newblock \bibinfo{journal}{\emph{SIGARCH Comput. Archit. News}} \bibinfo{volume}{39}, \bibinfo{number}{4} (\bibinfo{date}{Dec.} \bibinfo{year}{2011}), \bibinfo{pages}{98–99}.
\newblock
\showISSN{0163-5964}
\urldef\tempurl%
\url{https://doi.org/10.1145/2082156.2082183}
\showDOI{\tempurl}


\bibitem[Gholami(2021)]%
        {comp_bw_scaling}
\bibfield{author}{\bibinfo{person}{Amir Gholami}.} \bibinfo{year}{2021}\natexlab{}.
\newblock \bibinfo{title}{{AI and Memory Wall}}.
\newblock
\newblock


\bibitem[Gutierrez et~al\mbox{.}(2018)]%
        {GutierrezBeckmann2018-gem5GPU}
\bibfield{author}{\bibinfo{person}{Anthony Gutierrez}, \bibinfo{person}{Bradford~M. Beckmann}, \bibinfo{person}{Alexandru Dutu}, \bibinfo{person}{Joseph Gross}, \bibinfo{person}{Michael LeBeane}, \bibinfo{person}{John Kalamatianos}, \bibinfo{person}{Onur Kayiran}, \bibinfo{person}{Matthew Poremba}, \bibinfo{person}{Brandon Potter}, \bibinfo{person}{Sooraj Puthoor}, \bibinfo{person}{Matthew~D. Sinclair}, \bibinfo{person}{Michael Wyse}, \bibinfo{person}{Jieming Yin}, \bibinfo{person}{Xianwei Zhang}, \bibinfo{person}{Akshay Jain}, {and} \bibinfo{person}{Timothy Rogers}.} \bibinfo{year}{2018}\natexlab{}.
\newblock \showarticletitle{{Lost in Abstraction: Pitfalls of Analyzing GPUs at the Intermediate Language Level}}. In \bibinfo{booktitle}{\emph{{24th IEEE International Symposium on High Performance Computer Architecture}}} \emph{(\bibinfo{series}{HPCA})}. \bibinfo{publisher}{IEEE Computer Society}, \bibinfo{address}{Los Alamitos, CA, USA}, \bibinfo{pages}{608--619}.
\newblock
\showISSN{2378-203X}
\urldef\tempurl%
\url{https://doi.org/10.1109/HPCA.2018.00058}
\showDOI{\tempurl}


\bibitem[Hassan~Awadalla et~al\mbox{.}(2018)]%
        {hassan2018achieving}
\bibfield{author}{\bibinfo{person}{Hany Hassan~Awadalla}, \bibinfo{person}{Anthony Aue}, \bibinfo{person}{Chang Chen}, \bibinfo{person}{Vishal Chowdhary}, \bibinfo{person}{Jonathan Clark}, \bibinfo{person}{Christian Federmann}, \bibinfo{person}{Xuedong Huang}, \bibinfo{person}{Marcin Junczys-Dowmunt}, \bibinfo{person}{Will Lewis}, \bibinfo{person}{Mu Li}, \bibinfo{person}{Shujie Liu}, \bibinfo{person}{Tie-Yan Liu}, \bibinfo{person}{Renqian Luo}, \bibinfo{person}{Arul Menezes}, \bibinfo{person}{Tao Qin}, \bibinfo{person}{Frank Seide}, \bibinfo{person}{Xu Tan}, \bibinfo{person}{Fei Tian}, \bibinfo{person}{Lijun Wu}, \bibinfo{person}{Shuangzhi Wu}, \bibinfo{person}{Yingce Xia}, \bibinfo{person}{Dongdong Zhang}, \bibinfo{person}{Zhirui Zhang}, {and} \bibinfo{person}{Ming Zhou}.} \bibinfo{year}{2018}\natexlab{}.
\newblock \showarticletitle{{Achieving Human Parity on Automatic Chinese to English News Translation}}.
\newblock \bibinfo{journal}{\emph{arXiv preprint arXiv:1803.05567}} (\bibinfo{date}{March} \bibinfo{year}{2018}), \bibinfo{numpages}{25}~pages.
\newblock
\showeprint[arxiv]{1803.05567}~[cs.CL]


\bibitem[He et~al\mbox{.}(2015)]%
        {Resnet}
\bibfield{author}{\bibinfo{person}{Kaiming He}, \bibinfo{person}{Xiangyu Zhang}, \bibinfo{person}{Shaoqing Ren}, {and} \bibinfo{person}{Jian Sun}.} \bibinfo{year}{2015}\natexlab{}.
\newblock \showarticletitle{{Deep Residual Learning for Image Recognition}}.
\newblock \bibinfo{journal}{\emph{CoRR}}  \bibinfo{volume}{abs/1512.03385} (\bibinfo{year}{2015}), \bibinfo{numpages}{12}~pages.
\newblock
\showeprint[arxiv]{1512.03385}
\urldef\tempurl%
\url{http://arxiv.org/abs/1512.03385}
\showURL{%
\tempurl}


\bibitem[Huang et~al\mbox{.}(2019)]%
        {HuangCheng2019-gpipe}
\bibfield{author}{\bibinfo{person}{Yanping Huang}, \bibinfo{person}{Youlong Cheng}, \bibinfo{person}{Ankur Bapna}, \bibinfo{person}{Orhan Firat}, \bibinfo{person}{Mia~Xu Chen}, \bibinfo{person}{Dehao Chen}, \bibinfo{person}{HyoukJoong Lee}, \bibinfo{person}{Jiquan Ngiam}, \bibinfo{person}{Quoc~V. Le}, \bibinfo{person}{Yonghui Wu}, {and} \bibinfo{person}{Zhifeng Chen}.} \bibinfo{year}{2019}\natexlab{}.
\newblock \showarticletitle{{GPipe: Efficient Training of Giant Neural Networks using Pipeline Parallelism}}. In \bibinfo{booktitle}{\emph{{Proceedings of the 33rd International Conference on Neural Information Processing Systems}}} \emph{(\bibinfo{series}{{NeurIPS}}, Vol.~\bibinfo{volume}{32})}. \bibinfo{publisher}{Curran Associates Inc.}, \bibinfo{address}{Red Hook, NY, USA}, Article \bibinfo{articleno}{10}, \bibinfo{numpages}{10}~pages.
\newblock


\bibitem[Hwang et~al\mbox{.}(2023)]%
        {hwang2023ark}
\bibfield{author}{\bibinfo{person}{Changho Hwang}, \bibinfo{person}{KyoungSoo Park}, \bibinfo{person}{Ran Shu}, \bibinfo{person}{Xinyuan Qu}, \bibinfo{person}{Peng Cheng}, {and} \bibinfo{person}{Yongqiang Xiong}.} \bibinfo{year}{2023}\natexlab{}.
\newblock \showarticletitle{{ARK: GPU-driven Code Execution for Distributed Deep Learning}}. In \bibinfo{booktitle}{\emph{{20th USENIX Symposium on Networked Systems Design and Implementation}}} \emph{(\bibinfo{series}{NSDI})}. \bibinfo{publisher}{USENIX Association}, \bibinfo{address}{Boston, MA}, \bibinfo{pages}{87--101}.
\newblock
\showISBNx{978-1-939133-33-5}
\urldef\tempurl%
\url{https://www.usenix.org/conference/nsdi23/presentation/hwang}
\showURL{%
\tempurl}


\bibitem[Jangda et~al\mbox{.}(2022)]%
        {jangda2022breaking}
\bibfield{author}{\bibinfo{person}{Abhinav Jangda}, \bibinfo{person}{Jun Huang}, \bibinfo{person}{Guodong Liu}, \bibinfo{person}{Amir Hossein~Nodehi Sabet}, \bibinfo{person}{Saeed Maleki}, \bibinfo{person}{Youshan Miao}, \bibinfo{person}{Madanlal Musuvathi}, \bibinfo{person}{Todd Mytkowicz}, {and} \bibinfo{person}{Olli Saarikivi}.} \bibinfo{year}{2022}\natexlab{}.
\newblock \showarticletitle{{Breaking the Computation and Communication Abstraction Barrier in Distributed Machine Learning Workloads}}. In \bibinfo{booktitle}{\emph{{Proceedings of the 27th ACM International Conference on Architectural Support for Programming Languages and Operating Systems}}} \emph{(\bibinfo{series}{ASPLOS})}. \bibinfo{publisher}{Association for Computing Machinery}, \bibinfo{address}{New York, NY, USA}, \bibinfo{pages}{402--416}.
\newblock
\showISBNx{9781450392051}
\urldef\tempurl%
\url{https://doi.org/10.1145/3503222.3507778}
\showDOI{\tempurl}


\bibitem[Jeaugey(2022)]%
        {nccl-tree-ring}
\bibfield{author}{\bibinfo{person}{Sylvain Jeaugey}.} \bibinfo{year}{2022}\natexlab{}.
\newblock \bibinfo{title}{{How is tree reduction implemented?}}
\newblock \bibinfo{howpublished}{\url{https://github.com/NVIDIA/nccl/issues/545\#issuecomment-1006361565}}.
\newblock


\bibitem[Jog et~al\mbox{.}(2014)]%
        {jog2014application}
\bibfield{author}{\bibinfo{person}{Adwait Jog}, \bibinfo{person}{Evgeny Bolotin}, \bibinfo{person}{Zvika Guz}, \bibinfo{person}{Mike Parker}, \bibinfo{person}{Stephen~W Keckler}, \bibinfo{person}{Mahmut~T Kandemir}, {and} \bibinfo{person}{Chita~R Das}.} \bibinfo{year}{2014}\natexlab{}.
\newblock \showarticletitle{{Application-aware Memory System for Fair and Efficient Execution of Concurrent GPGPU Applications}}. In \bibinfo{booktitle}{\emph{{Proceedings of Workshop on General Purpose Processing using GPUs}}} \emph{(\bibinfo{series}{GPGPU})}. \bibinfo{publisher}{Association for Computing Machinery}, \bibinfo{address}{New York, NY, USA}, \bibinfo{pages}{1--8}.
\newblock
\showISBNx{9781450327664}
\urldef\tempurl%
\url{https://doi.org/10.1145/2588768.2576780}
\showDOI{\tempurl}


\bibitem[Jog et~al\mbox{.}(2016)]%
        {jog2016exploiting}
\bibfield{author}{\bibinfo{person}{Adwait Jog}, \bibinfo{person}{Onur Kayiran}, \bibinfo{person}{Ashutosh Pattnaik}, \bibinfo{person}{Mahmut~T Kandemir}, \bibinfo{person}{Onur Mutlu}, \bibinfo{person}{Ravishankar Iyer}, {and} \bibinfo{person}{Chita~R Das}.} \bibinfo{year}{2016}\natexlab{}.
\newblock \showarticletitle{{Exploiting Core Criticality for Enhanced GPU Performance}}. In \bibinfo{booktitle}{\emph{{Proceedings of the 2016 ACM SIGMETRICS International Conference on Measurement and Modeling of Computer Science}}}. \bibinfo{publisher}{Association for Computing Machinery}, \bibinfo{address}{New York, NY, USA}, \bibinfo{pages}{351--363}.
\newblock
\showISBNx{9781450342667}
\urldef\tempurl%
\url{https://doi.org/10.1145/2896377.2901468}
\showDOI{\tempurl}


\bibitem[Jouppi et~al\mbox{.}(2021)]%
        {JouppiYoon2021-tpuv4}
\bibfield{author}{\bibinfo{person}{Norman~P. Jouppi}, \bibinfo{person}{Doe~Hyun Yoon}, \bibinfo{person}{Matthew Ashcraft}, \bibinfo{person}{Mark Gottscho}, \bibinfo{person}{Thomas~B. Jablin}, \bibinfo{person}{George Kurian}, \bibinfo{person}{James Laudon}, \bibinfo{person}{Sheng Li}, \bibinfo{person}{Peter Ma}, \bibinfo{person}{Xiaoyu Ma}, \bibinfo{person}{Nishant Patil}, \bibinfo{person}{Sushma Prasad}, \bibinfo{person}{Clifford Young}, \bibinfo{person}{Zongwei Zhou}, {and} \bibinfo{person}{David Patterson}.} \bibinfo{year}{2021}\natexlab{}.
\newblock \showarticletitle{{Ten Lessons from Three Generations Shaped Google's TPUv4i}}. In \bibinfo{booktitle}{\emph{{Proceedings of the 48th Annual International Symposium on Computer Architecture}}} (Virtual Event, Spain) \emph{(\bibinfo{series}{ISCA})}. \bibinfo{publisher}{IEEE Press}, \bibinfo{address}{Piscataway, NJ, USA}, \bibinfo{pages}{1–14}.
\newblock
\showISBNx{9781450390866}
\urldef\tempurl%
\url{https://doi.org/10.1109/ISCA52012.2021.00010}
\showDOI{\tempurl}


\bibitem[Kerr et~al\mbox{.}(2017)]%
        {cutlass_nvidia}
\bibfield{author}{\bibinfo{person}{Andrew Kerr}, \bibinfo{person}{Duane Merrill}, \bibinfo{person}{Julien Demouth}, {and} \bibinfo{person}{John Tran}.} \bibinfo{year}{2017}\natexlab{}.
\newblock \bibinfo{title}{{cuTLASS: Fast linear algebra in CUDA C++}}.
\newblock
\newblock


\bibitem[Khairy et~al\mbox{.}(2018)]%
        {KhairyJain2018-voltaGPGPUSim}
\bibfield{author}{\bibinfo{person}{Mahmoud Khairy}, \bibinfo{person}{Akshay Jain}, \bibinfo{person}{Tor~M. Aamodt}, {and} \bibinfo{person}{Timothy~G. Rogers}.} \bibinfo{year}{2018}\natexlab{}.
\newblock \showarticletitle{{Exploring Modern GPU Memory System Design Challenges through Accurate Modeling}}.
\newblock \bibinfo{journal}{\emph{CoRR}}  \bibinfo{volume}{abs/1810.07269} (\bibinfo{year}{2018}), \bibinfo{numpages}{10}~pages.
\newblock
\showeprint[arxiv]{1810.07269}
\urldef\tempurl%
\url{http://arxiv.org/abs/1810.07269}
\showURL{%
\tempurl}


\bibitem[Khairy et~al\mbox{.}(2020a)]%
        {KhairyNikiforov2020-ladm}
\bibfield{author}{\bibinfo{person}{Mahmoud Khairy}, \bibinfo{person}{Vadim Nikiforov}, \bibinfo{person}{David Nellans}, {and} \bibinfo{person}{Timothy~G. Rogers}.} \bibinfo{year}{2020}\natexlab{a}.
\newblock \showarticletitle{{Locality-Centric Data and Threadblock Management for Massive GPUs}}. In \bibinfo{booktitle}{\emph{{53rd Annual IEEE/ACM International Symposium on Microarchitecture}}} \emph{(\bibinfo{series}{MICRO})}. \bibinfo{publisher}{IEEE Computer Society}, \bibinfo{address}{Los Alamitos, CA, USA}, \bibinfo{pages}{1022--1036}.
\newblock
\urldef\tempurl%
\url{https://doi.org/10.1109/MICRO50266.2020.00086}
\showDOI{\tempurl}


\bibitem[Khairy et~al\mbox{.}(2020b)]%
        {KhairyShen2021-accelSim}
\bibfield{author}{\bibinfo{person}{Mahmoud Khairy}, \bibinfo{person}{Zhesheng Shen}, \bibinfo{person}{Tor~M. Aamodt}, {and} \bibinfo{person}{Timothy~G. Rogers}.} \bibinfo{year}{2020}\natexlab{b}.
\newblock \showarticletitle{{Accel-Sim: An Extensible Simulation Framework for Validated GPU Modeling}}. In \bibinfo{booktitle}{\emph{{ACM/IEEE 47th Annual International Symposium on Computer Architecture}}} \emph{(\bibinfo{series}{ISCA})}. \bibinfo{publisher}{IEEE Press}, \bibinfo{address}{Piscataway, NJ, USA}, \bibinfo{pages}{473--486}.
\newblock
\showISBNx{9781728146614}
\urldef\tempurl%
\url{https://doi.org/10.1109/ISCA45697.2020.00047}
\showDOI{\tempurl}


\bibitem[Kim et~al\mbox{.}(2021b)]%
        {KimPark2021-gradPim}
\bibfield{author}{\bibinfo{person}{Heesu Kim}, \bibinfo{person}{Hanmin Park}, \bibinfo{person}{Taehyun Kim}, \bibinfo{person}{Kwanheum cho}, \bibinfo{person}{Eojin Lee}, \bibinfo{person}{Soojung Ryu}, \bibinfo{person}{Hyuk-Jae Lee}, \bibinfo{person}{Kiyoung Choi}, {and} \bibinfo{person}{Jinho Lee}.} \bibinfo{year}{2021}\natexlab{b}.
\newblock \showarticletitle{{GradPIM: A Practical Processing-in-DRAM Architecture for Gradient Descent}}. In \bibinfo{booktitle}{\emph{{27th IEEE International Symposium on High-Performance Computer Architecture}}} \emph{(\bibinfo{series}{HPCA})}. \bibinfo{publisher}{IEEE Computer Society}, \bibinfo{address}{Washington, DC, USA}, \bibinfo{numpages}{14}~pages.
\newblock


\bibitem[Kim et~al\mbox{.}(2021a)]%
        {Kim21Expertllsm}
\bibfield{author}{\bibinfo{person}{Young~Jin Kim}, \bibinfo{person}{Ammar~Ahmad Awan}, \bibinfo{person}{Alexandre Muzio}, \bibinfo{person}{Andres Felipe~Cruz Salinas}, \bibinfo{person}{Liyang Lu}, \bibinfo{person}{Amr Hendy}, \bibinfo{person}{Samyam Rajbhandari}, \bibinfo{person}{Yuxiong He}, {and} \bibinfo{person}{Hany~Hassan Awadalla}.} \bibinfo{year}{2021}\natexlab{a}.
\newblock \bibinfo{title}{{Scalable and Efficient MoE Training for Multitask Multilingual Models}}.
\newblock
\newblock
\urldef\tempurl%
\url{https://doi.org/10.48550/ARXIV.2109.10465}
\showDOI{\tempurl}


\bibitem[Klenk et~al\mbox{.}(2020)]%
        {klenk2020network}
\bibfield{author}{\bibinfo{person}{Benjamin Klenk}, \bibinfo{person}{Nan Jiang}, \bibinfo{person}{Greg Thorson}, {and} \bibinfo{person}{Larry Dennison}.} \bibinfo{year}{2020}\natexlab{}.
\newblock \showarticletitle{{An In-Network Architecture for Accelerating Shared-Memory Multiprocessor Collectives}}. In \bibinfo{booktitle}{\emph{{ACM/IEEE 47th Annual International Symposium on Computer Architecture}}} \emph{(\bibinfo{series}{ISCA})}. IEEE, \bibinfo{publisher}{IEEE Computer Society}, \bibinfo{address}{Washington, DC, USA}, \bibinfo{pages}{996--1009}.
\newblock


\bibitem[Krizhevsky et~al\mbox{.}(2012)]%
        {Krizhevsky12-alexnet}
\bibfield{author}{\bibinfo{person}{Alex Krizhevsky}, \bibinfo{person}{Ilya Sutskever}, {and} \bibinfo{person}{Geoffrey~E. Hinton}.} \bibinfo{year}{2012}\natexlab{}.
\newblock \showarticletitle{{ImageNet Classification with Deep Convolutional Neural Networks}}. In \bibinfo{booktitle}{\emph{{Proceedings of the 25th International Conference on Neural Information Processing Systems - Volume 1}}} (Lake Tahoe, Nevada) \emph{(\bibinfo{series}{NIPS'12})}. \bibinfo{publisher}{Curran Associates Inc.}, \bibinfo{address}{USA}, \bibinfo{pages}{1097--1105}.
\newblock
\urldef\tempurl%
\url{http://dl.acm.org/citation.cfm?id=2999134.2999257}
\showURL{%
\tempurl}


\bibitem[Lee et~al\mbox{.}(2021)]%
        {lee2021hardware}
\bibfield{author}{\bibinfo{person}{Sukhan Lee}, \bibinfo{person}{Shin-haeng Kang}, \bibinfo{person}{Jaehoon Lee}, \bibinfo{person}{Hyeonsu Kim}, \bibinfo{person}{Eojin Lee}, \bibinfo{person}{Seungwoo Seo}, \bibinfo{person}{Hosang Yoon}, \bibinfo{person}{Seungwon Lee}, \bibinfo{person}{Kyounghwan Lim}, \bibinfo{person}{Hyunsung Shin}, \bibinfo{person}{Jinhyun Kim}, \bibinfo{person}{O Seongil}, \bibinfo{person}{Anand Iyer}, \bibinfo{person}{David Wang}, \bibinfo{person}{Kyomin Sohn}, {and} \bibinfo{person}{Nam~Sung Kim}.} \bibinfo{year}{2021}\natexlab{}.
\newblock \showarticletitle{{Hardware Architecture and Software Stack for PIM Based on Commercial DRAM Technology: Industrial Product}}. In \bibinfo{booktitle}{\emph{{ACM/IEEE 48th Annual International Symposium on Computer Architecture}}} \emph{(\bibinfo{series}{ISCA})}. \bibinfo{publisher}{IEEE Press}, \bibinfo{address}{Piscataway, NJ, USA}, \bibinfo{pages}{43--56}.
\newblock
\showISBNx{9781450390866}
\urldef\tempurl%
\url{https://doi.org/10.1109/ISCA52012.2021.00013}
\showDOI{\tempurl}


\bibitem[Lew et~al\mbox{.}(2019)]%
        {lew19}
\bibfield{author}{\bibinfo{person}{Jonathan Lew}, \bibinfo{person}{Deval~A Shah}, \bibinfo{person}{Suchita Pati}, \bibinfo{person}{Shaylin Cattell}, \bibinfo{person}{Mengchi Zhang}, \bibinfo{person}{Amruth Sandhupatla}, \bibinfo{person}{Christopher Ng}, \bibinfo{person}{Negar Goli}, \bibinfo{person}{Matthew~D Sinclair}, \bibinfo{person}{Timothy~G Rogers}, {and} \bibinfo{person}{Tor Aamodt}.} \bibinfo{year}{2019}\natexlab{}.
\newblock \showarticletitle{{Analyzing Machine Learning Workloads Using a Detailed GPU Simulator}}. In \bibinfo{booktitle}{\emph{{IEEE International Symposium on Performance Analysis of Systems and Software}}} \emph{(\bibinfo{series}{ISPASS})}. IEEE, \bibinfo{publisher}{IEEE Computer Society}, \bibinfo{address}{Washington, DC, USA}, \bibinfo{pages}{151--152}.
\newblock


\bibitem[Lin et~al\mbox{.}(2014)]%
        {LinChen2013-nwInNW}
\bibfield{author}{\bibinfo{person}{Min Lin}, \bibinfo{person}{Qiang Chen}, {and} \bibinfo{person}{Shuicheng Yan}.} \bibinfo{year}{2014}\natexlab{}.
\newblock \showarticletitle{{Network In Network}}. In \bibinfo{booktitle}{\emph{{2nd International Conference on Learning Representations}}} \emph{(\bibinfo{series}{ICLR})}, \bibfield{editor}{\bibinfo{person}{Yoshua Bengio} {and} \bibinfo{person}{Yann LeCun}} (Eds.). \bibinfo{publisher}{OpenReview.net}, \bibinfo{numpages}{10}~pages.
\newblock
\urldef\tempurl%
\url{http://arxiv.org/abs/1312.4400}
\showURL{%
\tempurl}


\bibitem[Lin et~al\mbox{.}(2018)]%
        {LinZhang2018-avArch}
\bibfield{author}{\bibinfo{person}{Shih-Chieh Lin}, \bibinfo{person}{Yunqi Zhang}, \bibinfo{person}{Chang-Hong Hsu}, \bibinfo{person}{Matt Skach}, \bibinfo{person}{Md~E. Haque}, \bibinfo{person}{Lingjia Tang}, {and} \bibinfo{person}{Jason Mars}.} \bibinfo{year}{2018}\natexlab{}.
\newblock \showarticletitle{{The Architectural Implications of Autonomous Driving: Constraints and Acceleration}}. In \bibinfo{booktitle}{\emph{{Proceedings of the Twenty-Third International Conference on Architectural Support for Programming Languages and Operating Systems}}} (Williamsburg, VA, USA) \emph{(\bibinfo{series}{ASPLOS})}. \bibinfo{publisher}{ACM}, \bibinfo{address}{New York, NY, USA}, \bibinfo{pages}{751--766}.
\newblock
\showISBNx{978-1-4503-4911-6}
\urldef\tempurl%
\url{https://doi.org/10.1145/3173162.3173191}
\showDOI{\tempurl}


\bibitem[Lustig and Martonosi(2013)]%
        {lustig-full-empty}
\bibfield{author}{\bibinfo{person}{Daniel Lustig} {and} \bibinfo{person}{Margaret Martonosi}.} \bibinfo{year}{2013}\natexlab{}.
\newblock \showarticletitle{{Reducing GPU Offload Latency via Fine-Grained CPU-GPU Synchronization}}. In \bibinfo{booktitle}{\emph{{Proceedings of the 19th International Symposium on High Performance Computer Architecture}}} \emph{(\bibinfo{series}{HPCA})}. \bibinfo{publisher}{IEEE Computer Society}, \bibinfo{address}{USA}, \bibinfo{pages}{354–365}.
\newblock
\showISBNx{9781467355858}
\urldef\tempurl%
\url{https://doi.org/10.1109/HPCA.2013.6522332}
\showDOI{\tempurl}


\bibitem[Mahajan et~al\mbox{.}(2023)]%
        {mahajan2023better}
\bibfield{author}{\bibinfo{person}{Kshiteej Mahajan}, \bibinfo{person}{Ching-Hsiang Chu}, \bibinfo{person}{Srinivas Sridharan}, {and} \bibinfo{person}{Aditya Akella}.} \bibinfo{year}{2023}\natexlab{}.
\newblock \showarticletitle{{Better Together: Jointly Optimizing ML Collective Scheduling and Execution Planning using SYNDICATE}}. In \bibinfo{booktitle}{\emph{{20th USENIX Symposium on Networked Systems Design and Implementation}}} \emph{(\bibinfo{series}{NSDI})}. \bibinfo{publisher}{USENIX Association}, \bibinfo{address}{Boston, MA}, \bibinfo{pages}{809--824}.
\newblock
\showISBNx{978-1-939133-33-5}
\urldef\tempurl%
\url{https://www.usenix.org/conference/nsdi23/presentation/mahajan}
\showURL{%
\tempurl}


\bibitem[Mattson et~al\mbox{.}(2019)]%
        {MattsonCheng2019-mlperfTrain}
\bibfield{author}{\bibinfo{person}{Peter Mattson}, \bibinfo{person}{Christine Cheng}, \bibinfo{person}{Cody Coleman}, \bibinfo{person}{Greg Diamos}, \bibinfo{person}{Paulius Micikevicius}, \bibinfo{person}{David~A. Patterson}, \bibinfo{person}{Hanlin Tang}, \bibinfo{person}{Gu{-}Yeon Wei}, \bibinfo{person}{Peter Bailis}, \bibinfo{person}{Victor Bittorf}, \bibinfo{person}{David Brooks}, \bibinfo{person}{Dehao Chen}, \bibinfo{person}{Debojyoti Dutta}, \bibinfo{person}{Udit Gupta}, \bibinfo{person}{Kim~M. Hazelwood}, \bibinfo{person}{Andrew Hock}, \bibinfo{person}{Xinyuan Huang}, \bibinfo{person}{Bill Jia}, \bibinfo{person}{Daniel Kang}, \bibinfo{person}{David Kanter}, \bibinfo{person}{Naveen Kumar}, \bibinfo{person}{Jeffery Liao}, \bibinfo{person}{Guokai Ma}, \bibinfo{person}{Deepak Narayanan}, \bibinfo{person}{Tayo Oguntebi}, \bibinfo{person}{Gennady Pekhimenko}, \bibinfo{person}{Lillian Pentecost}, \bibinfo{person}{Vijay~Janapa Reddi}, \bibinfo{person}{Taylor Robie}, \bibinfo{person}{Tom~St. John},
  \bibinfo{person}{Carole{-}Jean Wu}, \bibinfo{person}{Lingjie Xu}, \bibinfo{person}{Cliff Young}, {and} \bibinfo{person}{Matei Zaharia}.} \bibinfo{year}{2019}\natexlab{}.
\newblock \showarticletitle{{MLPerf Training Benchmark}}.
\newblock \bibinfo{journal}{\emph{CoRR}}  \bibinfo{volume}{abs/1910.01500} (\bibinfo{year}{2019}), \bibinfo{numpages}{14}~pages.
\newblock
\showeprint[arxiv]{1910.01500}
\urldef\tempurl%
\url{http://arxiv.org/abs/1910.01500}
\showURL{%
\tempurl}


\bibitem[Micikevicius et~al\mbox{.}(2018)]%
        {micikevicius2018mixed}
\bibfield{author}{\bibinfo{person}{Paulius Micikevicius}, \bibinfo{person}{Sharan Narang}, \bibinfo{person}{Jonah Alben}, \bibinfo{person}{Gregory Diamos}, \bibinfo{person}{Erich Elsen}, \bibinfo{person}{David Garcia}, \bibinfo{person}{Boris Ginsburg}, \bibinfo{person}{Michael Houston}, \bibinfo{person}{Oleksii Kuchaiev}, \bibinfo{person}{Ganesh Venkatesh}, {and} \bibinfo{person}{Hao Wu}.} \bibinfo{year}{2018}\natexlab{}.
\newblock \bibinfo{title}{{Mixed Precision Training}}.
\newblock
\newblock
\showeprint[arxiv]{1710.03740}~[cs.AI]
\urldef\tempurl%
\url{http://arxiv.org/abs/1710.03740}
\showURL{%
\tempurl}


\bibitem[Micikevicius et~al\mbox{.}(2022)]%
        {IntelFP822}
\bibfield{author}{\bibinfo{person}{Paulius Micikevicius}, \bibinfo{person}{Dusan Stosic}, \bibinfo{person}{Neil Burgess}, \bibinfo{person}{Marius Cornea}, \bibinfo{person}{Pradeep Dubey}, \bibinfo{person}{Richard Grisenthwaite}, \bibinfo{person}{Sangwon Ha}, \bibinfo{person}{Alexander Heinecke}, \bibinfo{person}{Patrick Judd}, \bibinfo{person}{John Kamalu}, \bibinfo{person}{Naveen Mellempudi}, \bibinfo{person}{Stuart Oberman}, \bibinfo{person}{Mohammad Shoeybi}, \bibinfo{person}{Michael Siu}, {and} \bibinfo{person}{Hao Wu}.} \bibinfo{year}{2022}\natexlab{}.
\newblock \showarticletitle{{FP8 Formats for Deep Learning}}.
\newblock \bibinfo{journal}{\emph{CoRR}}  \bibinfo{volume}{abs/2209.05433} (\bibinfo{year}{2022}), \bibinfo{numpages}{9}~pages.
\newblock
\urldef\tempurl%
\url{https://doi.org/10.48550/ARXIV.2209.05433}
\showDOI{\tempurl}
\showeprint[arXiv]{2209.05433}


\bibitem[Microsoft(2020)]%
        {Microsoft2020-tnlg}
\bibfield{author}{\bibinfo{person}{Microsoft}.} \bibinfo{year}{2020}\natexlab{}.
\newblock \showarticletitle{{Turing-NLG: A 17-billion-parameter language model by Microsoft}}.
\newblock \bibinfo{journal}{\emph{Microsoft Research Blog}} \bibinfo{volume}{1}, \bibinfo{number}{8} (\bibinfo{year}{2020}), \bibinfo{numpages}{8}~pages.
\newblock
\urldef\tempurl%
\url{https://www.microsoft.com/en-us/research/blog/turing-nlg-a-17-billion-parameter-language-model-by-microsoft/}
\showURL{%
\tempurl}


\bibitem[MLPerf(2018)]%
        {mlperf}
\bibfield{author}{\bibinfo{person}{MLPerf}.} \bibinfo{year}{2018}\natexlab{}.
\newblock \bibinfo{title}{{MLPerf Benchmark Suite}}.
\newblock \bibinfo{howpublished}{\url{https://mlperf.org/}}.
\newblock


\bibitem[Moolchandani et~al\mbox{.}(2023)]%
        {moolchandani2023amped}
\bibfield{author}{\bibinfo{person}{Diksha Moolchandani}, \bibinfo{person}{Joyjit Kundu}, \bibinfo{person}{Frederik Ruelens}, \bibinfo{person}{Peter Vrancx}, \bibinfo{person}{Timon Evenblij}, {and} \bibinfo{person}{Manu Perumkunnil}.} \bibinfo{year}{2023}\natexlab{}.
\newblock \showarticletitle{{AMPeD: An Analytical Model for Performance in Distributed Training of Transformers}}. In \bibinfo{booktitle}{\emph{{IEEE International Symposium on Performance Analysis of Systems and Software}}} \emph{(\bibinfo{series}{ISPASS})}. \bibinfo{publisher}{IEEE Computer Society}, \bibinfo{address}{Los Alamitos, CA, USA}, \bibinfo{pages}{306--315}.
\newblock
\urldef\tempurl%
\url{https://doi.org/10.1109/ISPASS57527.2023.00037}
\showDOI{\tempurl}


\bibitem[Muthukrishnan et~al\mbox{.}(2021a)]%
        {muthukrishnan2021gps}
\bibfield{author}{\bibinfo{person}{Harini Muthukrishnan}, \bibinfo{person}{Daniel Lustig}, \bibinfo{person}{David Nellans}, {and} \bibinfo{person}{Thomas Wenisch}.} \bibinfo{year}{2021}\natexlab{a}.
\newblock \showarticletitle{GPS: A Global Publish-Subscribe Model for Multi-GPU Memory Management}. In \bibinfo{booktitle}{\emph{MICRO-54: 54th Annual IEEE/ACM International Symposium on Microarchitecture}} (Virtual Event, Greece) \emph{(\bibinfo{series}{MICRO '21})}. \bibinfo{publisher}{Association for Computing Machinery}, \bibinfo{address}{New York, NY, USA}, \bibinfo{pages}{46–58}.
\newblock
\showISBNx{9781450385572}
\urldef\tempurl%
\url{https://doi.org/10.1145/3466752.3480088}
\showDOI{\tempurl}


\bibitem[Muthukrishnan et~al\mbox{.}(2021b)]%
        {muthukrishnan2021efficient}
\bibfield{author}{\bibinfo{person}{Harini Muthukrishnan}, \bibinfo{person}{David Nellans}, \bibinfo{person}{Daniel Lustig}, \bibinfo{person}{Jeffrey~A Fessler}, {and} \bibinfo{person}{Thomas~F Wenisch}.} \bibinfo{year}{2021}\natexlab{b}.
\newblock \showarticletitle{{Efficient Multi-GPU Shared Memory via Automatic Optimization of Fine-grained Rransfers}}. In \bibinfo{booktitle}{\emph{{ACM/IEEE 48th Annual International Symposium on Computer Architecture}}} \emph{(\bibinfo{series}{ISCA})}. IEEE, \bibinfo{publisher}{IEEE Computer Society}, \bibinfo{address}{Washington, DC, USA}, \bibinfo{pages}{139--152}.
\newblock


\bibitem[Nai et~al\mbox{.}(2017)]%
        {nai2017graphpim}
\bibfield{author}{\bibinfo{person}{Lifeng Nai}, \bibinfo{person}{Ramyad Hadidi}, \bibinfo{person}{Jaewoong Sim}, \bibinfo{person}{Hyojong Kim}, \bibinfo{person}{Pranith Kumar}, {and} \bibinfo{person}{Hyesoon Kim}.} \bibinfo{year}{2017}\natexlab{}.
\newblock \showarticletitle{{GraphPIM: Enabling Instruction-level PIM Offloading in Graph Computing Frameworks}}. In \bibinfo{booktitle}{\emph{{IEEE International Symposium on High Performance Computer Architecture}}} \emph{(\bibinfo{series}{HPCA})}. IEEE, \bibinfo{publisher}{IEEE Computer Society}, \bibinfo{address}{Los Alamitos, CA, USA}, \bibinfo{pages}{457--468}.
\newblock
\urldef\tempurl%
\url{https://doi.org/10.1109/HPCA.2017.54}
\showDOI{\tempurl}


\bibitem[NVIDIA(2017)]%
        {dgx-v100}
\bibfield{author}{\bibinfo{person}{NVIDIA}.} \bibinfo{year}{2017}\natexlab{}.
\newblock \bibinfo{title}{{NVIDIA DGX-1 With Tesla V100 System Architecture}}.
\newblock \bibinfo{howpublished}{https://images.nvidia.com/content/pdf/dgx1-v100-system-architecture-whitepaper.pdf}.
\newblock


\bibitem[{NVIDIA}(2018)]%
        {v100}
\bibfield{author}{\bibinfo{person}{{NVIDIA}}.} \bibinfo{year}{2018}\natexlab{}.
\newblock \bibinfo{title}{{NVIDIA TESLA V100 GPU ACCELERATOR}}.
\newblock \bibinfo{howpublished}{\url{https://images.nvidia.com/content/technologies/volta/pdf/tesla-volta-v100-datasheet-letter-fnl-web.pdf}}.
\newblock


\bibitem[NVIDIA(2020)]%
        {nccl}
\bibfield{author}{\bibinfo{person}{NVIDIA}.} \bibinfo{year}{2020}\natexlab{}.
\newblock \bibinfo{title}{{NVIDIA NCCL}}.
\newblock
\newblock


\bibitem[{NVIDIA}(2021)]%
        {a100}
\bibfield{author}{\bibinfo{person}{{NVIDIA}}.} \bibinfo{year}{2021}\natexlab{}.
\newblock \bibinfo{title}{{NVIDIA A100 TENSOR CORE GPU}}.
\newblock \bibinfo{howpublished}{\url{https://www.nvidia.com/content/dam/en-zz/Solutions/Data-Center/a100/pdf/nvidia-a100-datasheet-us-nvidia-1758950-r4-web.pdf}}.
\newblock


\bibitem[{NVIDIA}(2022)]%
        {gpudirect}
\bibfield{author}{\bibinfo{person}{{NVIDIA}}.} \bibinfo{year}{2022}\natexlab{}.
\newblock \bibinfo{title}{{GPUDirect}}.
\newblock \bibinfo{howpublished}{"https://developer.nvidia. com/gpudirect"}.
\newblock


\bibitem[{NVIDIA}(2023a)]%
        {splitk}
\bibfield{author}{\bibinfo{person}{{NVIDIA}}.} \bibinfo{year}{2023}\natexlab{a}.
\newblock \bibinfo{title}{{Efficient GEMM in CUDA}}.
\newblock \bibinfo{howpublished}{\url{https://github.com/NVIDIA/cutlass/blob/main/media/docs/efficient_gemm.md\#parallelized-reductions}}.
\newblock


\bibitem[{NVIDIA}(2023b)]%
        {GH200}
\bibfield{author}{\bibinfo{person}{{NVIDIA}}.} \bibinfo{year}{2023}\natexlab{b}.
\newblock \bibinfo{title}{{NVIDIA Announces DGX GH200 AI Supercomputer}}.
\newblock \bibinfo{howpublished}{\url{https://nvidianews.nvidia.com/news/nvidia-grace-hopper-superchips-designed-for-accelerated-generative-ai-enter-full-production}}.
\newblock


\bibitem[{NVIDIA}(2023c)]%
        {h100}
\bibfield{author}{\bibinfo{person}{{NVIDIA}}.} \bibinfo{year}{2023}\natexlab{c}.
\newblock \bibinfo{title}{{NVIDIA H100 TENSOR CORE GPU}}.
\newblock \bibinfo{howpublished}{\url{https://resources.nvidia.com/en-us-tensor-core/nvidia-tensor-core-gpu-datasheet}}.
\newblock


\bibitem[{NVIDIA Corp.}(2016)]%
        {cublas}
\bibfield{author}{\bibinfo{person}{{NVIDIA Corp.}}} \bibinfo{year}{2016}\natexlab{}.
\newblock \bibinfo{title}{{NVIDIA cuBLAS}}.
\newblock \bibinfo{howpublished}{\url{https://developer.nvidia.com/cublas}}.
\newblock
\newblock
\shownote{Accessed August 6, 2016}.


\bibitem[Patel et~al\mbox{.}(2023)]%
        {patel2023splitwise}
\bibfield{author}{\bibinfo{person}{Pratyush Patel}, \bibinfo{person}{Esha Choukse}, \bibinfo{person}{Chaojie Zhang}, \bibinfo{person}{{\'I}{\~n}igo Goiri}, \bibinfo{person}{Aashaka Shah}, \bibinfo{person}{Saeed Maleki}, {and} \bibinfo{person}{Ricardo Bianchini}.} \bibinfo{year}{2023}\natexlab{}.
\newblock \showarticletitle{{Splitwise: Efficient Generative LLM Inference Using Phase Splitting}}.
\newblock \bibinfo{journal}{\emph{arXiv preprint arXiv:2311.18677}} (\bibinfo{year}{2023}), \bibinfo{numpages}{12}~pages.
\newblock
\showeprint[arxiv]{2311.18677}~[cs.AR]


\bibitem[Pati et~al\mbox{.}(2022)]%
        {PatiAga2021-demystifying}
\bibfield{author}{\bibinfo{person}{Suchita Pati}, \bibinfo{person}{Shaizeen Aga}, \bibinfo{person}{Nuwan Jayasena}, {and} \bibinfo{person}{Matthew~D. Sinclair}.} \bibinfo{year}{2022}\natexlab{}.
\newblock \showarticletitle{Demystifying BERT: System Design Implications}. In \bibinfo{booktitle}{\emph{2022 IEEE International Symposium on Workload Characterization (IISWC)}}. \bibinfo{publisher}{IEEE Computer Society}, \bibinfo{address}{Los Alamitos, CA, USA}, \bibinfo{pages}{296--309}.
\newblock
\urldef\tempurl%
\url{https://doi.org/10.1109/IISWC55918.2022.00033}
\showDOI{\tempurl}


\bibitem[Pati et~al\mbox{.}(2023)]%
        {pati2023computation}
\bibfield{author}{\bibinfo{person}{Suchita Pati}, \bibinfo{person}{Shaizeen Aga}, \bibinfo{person}{Nuwan Jayasena}, {and} \bibinfo{person}{Matthew~D. Sinclair}.} \bibinfo{year}{2023}\natexlab{}.
\newblock \showarticletitle{{Tale of Two Cs: Computation vs. Communication Scaling for Future Transformers on Future Hardware}}. In \bibinfo{booktitle}{\emph{{IEEE International Symposium on Workload Characterization}}} \emph{(\bibinfo{series}{IISWC})}. \bibinfo{publisher}{IEEE Computer Society}, \bibinfo{address}{Los Alamitos, CA, USA}, \bibinfo{pages}{140--153}.
\newblock
\urldef\tempurl%
\url{https://doi.org/10.1109/IISWC59245.2023.00026}
\showDOI{\tempurl}


\bibitem[Pattnaik et~al\mbox{.}(2019)]%
        {pattnaik2019opportunistic}
\bibfield{author}{\bibinfo{person}{Ashutosh Pattnaik}, \bibinfo{person}{Xulong Tang}, \bibinfo{person}{Onur Kayiran}, \bibinfo{person}{Adwait Jog}, \bibinfo{person}{Asit Mishra}, \bibinfo{person}{Mahmut~T Kandemir}, \bibinfo{person}{Anand Sivasubramaniam}, {and} \bibinfo{person}{Chita~R Das}.} \bibinfo{year}{2019}\natexlab{}.
\newblock \showarticletitle{{Opportunistic Computing in GPU Architectures}}. In \bibinfo{booktitle}{\emph{{Proceedings of the 46th International Symposium on Computer Architecture}}} \emph{(\bibinfo{series}{ISCA})}. \bibinfo{publisher}{Association for Computing Machinery}, \bibinfo{address}{New York, NY, USA}, \bibinfo{pages}{210--223}.
\newblock
\showISBNx{9781450366694}
\urldef\tempurl%
\url{https://doi.org/10.1145/3307650.3322212}
\showDOI{\tempurl}


\bibitem[Pawlowski(2011)]%
        {pawlowski2011hybrid}
\bibfield{author}{\bibinfo{person}{J~Thomas Pawlowski}.} \bibinfo{year}{2011}\natexlab{}.
\newblock \showarticletitle{{Hybrid Memory Cube (HMC)}}. In \bibinfo{booktitle}{\emph{{2011 IEEE Hot Chips 23 Symposium}}} \emph{(\bibinfo{series}{HotChips})}. IEEE, \bibinfo{publisher}{IEEE}, \bibinfo{address}{Piscataway, NJ, USA}, \bibinfo{pages}{1--24}.
\newblock


\bibitem[Punniyamurthy et~al\mbox{.}(2023)]%
        {punniyamurthy2023gpu}
\bibfield{author}{\bibinfo{person}{Kishore Punniyamurthy}, \bibinfo{person}{Bradford~M Beckmann}, {and} \bibinfo{person}{Khaled Hamidouche}.} \bibinfo{year}{2023}\natexlab{}.
\newblock \showarticletitle{{GPU-initiated Fine-grained Overlap of Collective Communication with Computation}}.
\newblock \bibinfo{journal}{\emph{arXiv preprint arXiv:2305.06942}} (\bibinfo{year}{2023}), \bibinfo{numpages}{13}~pages.
\newblock
\showeprint[arxiv]{2305.06942}~[cs.DC]


\bibitem[Radford et~al\mbox{.}(2019)]%
        {RadfordWu2019-gpt2}
\bibfield{author}{\bibinfo{person}{Alec Radford}, \bibinfo{person}{Jeffrey Wu}, \bibinfo{person}{Rewon Child}, \bibinfo{person}{David Luan}, \bibinfo{person}{Dario Amodei}, {and} \bibinfo{person}{Ilya Sutskever}.} \bibinfo{year}{2019}\natexlab{}.
\newblock \showarticletitle{{Language Models are Unsupervised Multitask Learners}}.
\newblock \bibinfo{journal}{\emph{OpenAI blog}} \bibinfo{volume}{1}, \bibinfo{number}{8} (\bibinfo{year}{2019}), \bibinfo{pages}{9}.
\newblock


\bibitem[Rajbhandari et~al\mbox{.}(2022)]%
        {rajbhandari2022deepspeed}
\bibfield{author}{\bibinfo{person}{Samyam Rajbhandari}, \bibinfo{person}{Conglong Li}, \bibinfo{person}{Zhewei Yao}, \bibinfo{person}{Minjia Zhang}, \bibinfo{person}{Reza~Yazdani Aminabadi}, \bibinfo{person}{Ammar~Ahmad Awan}, \bibinfo{person}{Jeff Rasley}, {and} \bibinfo{person}{Yuxiong He}.} \bibinfo{year}{2022}\natexlab{}.
\newblock \showarticletitle{{DeepSpeed-MOE: Advancing Mixture-of-Experts Inference and Training to Power Next-Generation AI Scale}}. In \bibinfo{booktitle}{\emph{{International Conference on Machine Learning}}} \emph{(\bibinfo{series}{ICML})}. PMLR, \bibinfo{publisher}{PMLR}, \bibinfo{pages}{18332--18346}.
\newblock


\bibitem[Rajbhandari et~al\mbox{.}(2021)]%
        {ZeRO-Infinity21}
\bibfield{author}{\bibinfo{person}{Samyam Rajbhandari}, \bibinfo{person}{Olatunji Ruwase}, \bibinfo{person}{Jeff Rasley}, \bibinfo{person}{Shaden Smith}, {and} \bibinfo{person}{Yuxiong He}.} \bibinfo{year}{2021}\natexlab{}.
\newblock \showarticletitle{{ZeRO-Infinity: Breaking the GPU Memory Wall for Extreme Scale Deep Learning}}. In \bibinfo{booktitle}{\emph{{Proceedings of the International Conference for High Performance Computing, Networking, Storage and Analysis}}} (St. Louis, Missouri) \emph{(\bibinfo{series}{SC '21})}. \bibinfo{publisher}{Association for Computing Machinery}, \bibinfo{address}{New York, NY, USA}, Article \bibinfo{articleno}{59}, \bibinfo{numpages}{14}~pages.
\newblock
\showISBNx{9781450384421}
\urldef\tempurl%
\url{https://doi.org/10.1145/3458817.3476205}
\showDOI{\tempurl}


\bibitem[Rashidi et~al\mbox{.}(2021)]%
        {rashidi2021enabling}
\bibfield{author}{\bibinfo{person}{Saeed Rashidi}, \bibinfo{person}{Matthew Denton}, \bibinfo{person}{Srinivas Sridharan}, \bibinfo{person}{Sudarshan Srinivasan}, \bibinfo{person}{Amoghavarsha Suresh}, \bibinfo{person}{Jade Nie}, {and} \bibinfo{person}{Tushar Krishna}.} \bibinfo{year}{2021}\natexlab{}.
\newblock \showarticletitle{{Enabling Compute-Communication Overlap in Distributed Deep Learning Training Platforms}}. In \bibinfo{booktitle}{\emph{{2021 ACM/IEEE 48th Annual International Symposium on Computer Architecture}}} \emph{(\bibinfo{series}{ISCA})}. IEEE, \bibinfo{publisher}{IEEE Press}, \bibinfo{address}{Piscataway, NJ, USA}, \bibinfo{pages}{540--553}.
\newblock
\showISBNx{9781450390866}
\urldef\tempurl%
\url{https://doi.org/10.1109/ISCA52012.2021.00049}
\showDOI{\tempurl}


\bibitem[{Reddi} et~al\mbox{.}(2020)]%
        {ReddiCheng2020-mlperfInfer}
\bibfield{author}{\bibinfo{person}{V.~J. {Reddi}}, \bibinfo{person}{C. {Cheng}}, \bibinfo{person}{D. {Kanter}}, \bibinfo{person}{P. {Mattson}}, \bibinfo{person}{G. {Schmuelling}}, \bibinfo{person}{C. {Wu}}, \bibinfo{person}{B. {Anderson}}, \bibinfo{person}{M. {Breughe}}, \bibinfo{person}{M. {Charlebois}}, \bibinfo{person}{W. {Chou}}, \bibinfo{person}{R. {Chukka}}, \bibinfo{person}{C. {Coleman}}, \bibinfo{person}{S. {Davis}}, \bibinfo{person}{P. {Deng}}, \bibinfo{person}{G. {Diamos}}, \bibinfo{person}{J. {Duke}}, \bibinfo{person}{D. {Fick}}, \bibinfo{person}{J.~S. {Gardner}}, \bibinfo{person}{I. {Hubara}}, \bibinfo{person}{S. {Idgunji}}, \bibinfo{person}{T.~B. {Jablin}}, \bibinfo{person}{J. {Jiao}}, \bibinfo{person}{T.~S. {John}}, \bibinfo{person}{P. {Kanwar}}, \bibinfo{person}{D. {Lee}}, \bibinfo{person}{J. {Liao}}, \bibinfo{person}{A. {Lokhmotov}}, \bibinfo{person}{F. {Massa}}, \bibinfo{person}{P. {Meng}}, \bibinfo{person}{P. {Micikevicius}}, \bibinfo{person}{C. {Osborne}}, \bibinfo{person}{G. {Pekhimenko}},
  \bibinfo{person}{A.~T.~R. {Rajan}}, \bibinfo{person}{D. {Sequeira}}, \bibinfo{person}{A. {Sirasao}}, \bibinfo{person}{F. {Sun}}, \bibinfo{person}{H. {Tang}}, \bibinfo{person}{M. {Thomson}}, \bibinfo{person}{F. {Wei}}, \bibinfo{person}{E. {Wu}}, \bibinfo{person}{L. {Xu}}, \bibinfo{person}{K. {Yamada}}, \bibinfo{person}{B. {Yu}}, \bibinfo{person}{G. {Yuan}}, \bibinfo{person}{A. {Zhong}}, \bibinfo{person}{P. {Zhang}}, {and} \bibinfo{person}{Y. {Zhou}}.} \bibinfo{year}{2020}\natexlab{}.
\newblock \showarticletitle{{MLPerf Inference Benchmark}}. In \bibinfo{booktitle}{\emph{{ACM/IEEE 47th Annual International Symposium on Computer Architecture}}} \emph{(\bibinfo{series}{ISCA})}. \bibinfo{publisher}{IEEE Press}, \bibinfo{address}{Washington, DC, USA}, \bibinfo{pages}{446--459}.
\newblock
\showISBNx{9781728146614}
\urldef\tempurl%
\url{https://doi.org/10.1109/ISCA45697.2020.00045}
\showDOI{\tempurl}


\bibitem[Reed et~al\mbox{.}(2022)]%
        {DeepMind-Gato22}
\bibfield{author}{\bibinfo{person}{Scott Reed}, \bibinfo{person}{Konrad Zolna}, \bibinfo{person}{Emilio Parisotto}, \bibinfo{person}{Sergio~Gomez Colmenarejo}, \bibinfo{person}{Alexander Novikov}, \bibinfo{person}{Gabriel Barth-Maron}, \bibinfo{person}{Mai Gimenez}, \bibinfo{person}{Yury Sulsky}, \bibinfo{person}{Jackie Kay}, \bibinfo{person}{Jost~Tobias Springenberg}, \bibinfo{person}{Tom Eccles}, \bibinfo{person}{Jake Bruce}, \bibinfo{person}{Ali Razavi}, \bibinfo{person}{Ashley Edwards}, \bibinfo{person}{Nicolas Heess}, \bibinfo{person}{Yutian Chen}, \bibinfo{person}{Raia Hadsell}, \bibinfo{person}{Oriol Vinyals}, \bibinfo{person}{Mahyar Bordbar}, {and} \bibinfo{person}{Nando de Freitas}.} \bibinfo{year}{2022}\natexlab{}.
\newblock \showarticletitle{{A Generalist Agent}}.
\newblock \bibinfo{journal}{\emph{{Transactions on Machine Learning Research}}}  \bibinfo{volume}{2022} (\bibinfo{year}{2022}), \bibinfo{numpages}{42}~pages.
\newblock
\urldef\tempurl%
\url{https://openreview.net/forum?id=1ikK0kHjvj}
\showURL{%
\tempurl}


\bibitem[Roarty and Sinclair(2020)]%
        {RoartySinclair2020-gem5GPU}
\bibfield{author}{\bibinfo{person}{Kyle Roarty} {and} \bibinfo{person}{Matthew~D. Sinclair}.} \bibinfo{year}{2020}\natexlab{}.
\newblock \showarticletitle{{Modeling Modern GPU Applications in gem5}}. In \bibinfo{booktitle}{\emph{{3rd gem5 Users' Workshop}}}. \bibinfo{numpages}{2}~pages.
\newblock


\bibitem[Rouhani et~al\mbox{.}(2020)]%
        {MSFT-BFP20}
\bibfield{author}{\bibinfo{person}{Bita Rouhani}, \bibinfo{person}{Daniel Lo}, \bibinfo{person}{Ritchie Zhao}, \bibinfo{person}{Ming Liu}, \bibinfo{person}{Jeremy Fowers}, \bibinfo{person}{Kalin Ovtcharov}, \bibinfo{person}{Anna Vinogradsky}, \bibinfo{person}{Sarah Massengill}, \bibinfo{person}{Lita Yang}, \bibinfo{person}{Ray Bittner}, \bibinfo{person}{Alessandro Forin}, \bibinfo{person}{Haishan Zhu}, \bibinfo{person}{Taesik Na}, \bibinfo{person}{Prerak Patel}, \bibinfo{person}{Shuai Che}, \bibinfo{person}{Lok~Chand Koppaka}, \bibinfo{person}{Xia Song}, \bibinfo{person}{Subhojit Som}, \bibinfo{person}{Kaustav Das}, \bibinfo{person}{Saurabh Tiwary}, \bibinfo{person}{Steve Reinhardt}, \bibinfo{person}{Sitaram Lanka}, \bibinfo{person}{Eric Chung}, {and} \bibinfo{person}{Doug Burger}.} \bibinfo{year}{2020}\natexlab{}.
\newblock \showarticletitle{{Pushing the Limits of Narrow Precision Inferencing at Cloud Scale with Microsoft Floating Point}}. In \bibinfo{booktitle}{\emph{{Proceedings of the 34th International Conference on Neural Information Processing Systems}}} (Vancouver, BC, Canada) \emph{(\bibinfo{series}{NeurIPS'20})}. \bibinfo{publisher}{Curran Associates Inc.}, \bibinfo{address}{Red Hook, NY, USA}, Article \bibinfo{articleno}{861}, \bibinfo{numpages}{11}~pages.
\newblock
\showISBNx{9781713829546}


\bibitem[Selvan and Kanwar(2022)]%
        {tpu_pod}
\bibfield{author}{\bibinfo{person}{Aarush Selvan} {and} \bibinfo{person}{Pankaj Kanwar}.} \bibinfo{year}{2022}\natexlab{}.
\newblock \bibinfo{title}{{Google showcases Cloud TPU v4 Pods for large model training}}.
\newblock \bibinfo{howpublished}{\url{https://cloud.google.com/blog/topics/tpus/google-showcases-cloud-tpu-v4-pods-for-large-model-training}}.
\newblock


\bibitem[Shah et~al\mbox{.}(2023)]%
        {shah2023taccl}
\bibfield{author}{\bibinfo{person}{Aashaka Shah}, \bibinfo{person}{Vijay Chidambaram}, \bibinfo{person}{Meghan Cowan}, \bibinfo{person}{Saeed Maleki}, \bibinfo{person}{Madan Musuvathi}, \bibinfo{person}{Todd Mytkowicz}, \bibinfo{person}{Jacob Nelson}, \bibinfo{person}{Olli Saarikivi}, {and} \bibinfo{person}{Rachee Singh}.} \bibinfo{year}{2023}\natexlab{}.
\newblock \showarticletitle{{TACCL: Guiding Collective Algorithm Synthesis using Communication Sketches}}. In \bibinfo{booktitle}{\emph{{20th USENIX Symposium on Networked Systems Design and Implementation}}} \emph{(\bibinfo{series}{NSDI})}. \bibinfo{publisher}{USENIX Association}, \bibinfo{address}{Boston, MA}, \bibinfo{pages}{593--612}.
\newblock
\showISBNx{978-1-939133-33-5}
\urldef\tempurl%
\url{https://www.usenix.org/conference/nsdi23/presentation/shah}
\showURL{%
\tempurl}


\bibitem[Shoeybi et~al\mbox{.}(2019)]%
        {ShoeybiPatwary2019-megatronlm}
\bibfield{author}{\bibinfo{person}{Mohammad Shoeybi}, \bibinfo{person}{Mostofa Patwary}, \bibinfo{person}{Raul Puri}, \bibinfo{person}{Patrick LeGresley}, \bibinfo{person}{Jared Casper}, {and} \bibinfo{person}{Bryan Catanzaro}.} \bibinfo{year}{2019}\natexlab{}.
\newblock \showarticletitle{{Megatron-LM: Training Multi-Billion Parameter Language Models Using Model Parallelism}}.
\newblock \bibinfo{journal}{\emph{CoRR}}  \bibinfo{volume}{abs/1909.08053} (\bibinfo{year}{2019}), \bibinfo{numpages}{9}~pages.
\newblock
\showeprint[arxiv]{1909.08053}~[cs.CL]
\urldef\tempurl%
\url{http://arxiv.org/abs/1909.08053}
\showURL{%
\tempurl}


\bibitem[Simonyan and Zisserman(2015)]%
        {SimonyanZisserman2014-imageRec}
\bibfield{author}{\bibinfo{person}{Karen Simonyan} {and} \bibinfo{person}{Andrew Zisserman}.} \bibinfo{year}{2015}\natexlab{}.
\newblock \showarticletitle{{Very Deep Convolutional Networks for Large-Scale Image Recognition}}. In \bibinfo{booktitle}{\emph{{3rd International Conference on Learning Representations}}} \emph{(\bibinfo{series}{ICLR})}, \bibfield{editor}{\bibinfo{person}{Yoshua Bengio} {and} \bibinfo{person}{Yann LeCun}} (Eds.). \bibinfo{publisher}{OpenReview.net}, \bibinfo{numpages}{14}~pages.
\newblock
\urldef\tempurl%
\url{http://arxiv.org/abs/1409.1556}
\showURL{%
\tempurl}


\bibitem[Smith et~al\mbox{.}(2022)]%
        {smith2022using}
\bibfield{author}{\bibinfo{person}{Shaden Smith}, \bibinfo{person}{Mostofa Patwary}, \bibinfo{person}{Brandon Norick}, \bibinfo{person}{Patrick LeGresley}, \bibinfo{person}{Samyam Rajbhandari}, \bibinfo{person}{Jared Casper}, \bibinfo{person}{Zhun Liu}, \bibinfo{person}{Shrimai Prabhumoye}, \bibinfo{person}{George Zerveas}, \bibinfo{person}{Vijay Korthikanti}, \bibinfo{person}{Elton Zhang}, \bibinfo{person}{Rewon Child}, \bibinfo{person}{Reza Yazdani~Aminabadi}, \bibinfo{person}{Julie Bernauer}, \bibinfo{person}{Xia Song}, \bibinfo{person}{Mohammad Shoeybi}, \bibinfo{person}{Yuxiong He}, \bibinfo{person}{Michael Houston}, \bibinfo{person}{Saurabh Tiwary}, {and} \bibinfo{person}{Bryan Catanzaro}.} \bibinfo{year}{2022}\natexlab{}.
\newblock \showarticletitle{{Using DeepSpeed and Megatron to Train Megatron-Turing NLG 530b, a Large-scale Generative Language Model}}.
\newblock \bibinfo{journal}{\emph{arXiv preprint arXiv:2201.11990}} (\bibinfo{year}{2022}), \bibinfo{numpages}{44}~pages.
\newblock
\showeprint[arxiv]{2201.11990}~[cs.CL]


\bibitem[Springer et~al\mbox{.}(2017)]%
        {SpringerWauligmann2017-fuseGPULang}
\bibfield{author}{\bibinfo{person}{Matthias Springer}, \bibinfo{person}{Peter Wauligmann}, {and} \bibinfo{person}{Hidehiko Masuhara}.} \bibinfo{year}{2017}\natexlab{}.
\newblock \showarticletitle{{Modular Array-Based GPU Computing in a Dynamically-Typed Language}}. In \bibinfo{booktitle}{\emph{{Proceedings of the 4th ACM SIGPLAN International Workshop on Libraries, Languages, and Compilers for Array Programming}}} (Barcelona, Spain) \emph{(\bibinfo{series}{ARRAY 2017})}. \bibinfo{publisher}{Association for Computing Machinery}, \bibinfo{address}{New York, NY, USA}, \bibinfo{pages}{48–55}.
\newblock
\showISBNx{9781450350693}
\urldef\tempurl%
\url{https://doi.org/10.1145/3091966.3091974}
\showDOI{\tempurl}


\bibitem[Szegedy et~al\mbox{.}(2015)]%
        {SzegedyLiu2015-deeperConv}
\bibfield{author}{\bibinfo{person}{Christian Szegedy}, \bibinfo{person}{Wei Liu}, \bibinfo{person}{Yangqing Jia}, \bibinfo{person}{Pierre Sermanet}, \bibinfo{person}{Scott Reed}, \bibinfo{person}{Dragomir Anguelov}, \bibinfo{person}{Dumitru Erhan}, \bibinfo{person}{Vincent Vanhoucke}, {and} \bibinfo{person}{Andrew Rabinovich}.} \bibinfo{year}{2015}\natexlab{}.
\newblock \showarticletitle{{Going Deeper with Convolutions}}. In \bibinfo{booktitle}{\emph{{Proceedings of the IEEE Conference on Computer Vision and Pattern Recognition}}} \emph{(\bibinfo{series}{CVPR})}. \bibinfo{publisher}{IEEE Press}, \bibinfo{address}{Piscataway, NJ, USA}, \bibinfo{pages}{1--9}.
\newblock


\bibitem[Szegedy et~al\mbox{.}(2016)]%
        {SzegedyVanhoucke2015-cv}
\bibfield{author}{\bibinfo{person}{Christian Szegedy}, \bibinfo{person}{Vincent Vanhoucke}, \bibinfo{person}{Sergey Ioffe}, \bibinfo{person}{Jonathon Shlens}, {and} \bibinfo{person}{Zbigniew Wojna}.} \bibinfo{year}{2016}\natexlab{}.
\newblock \showarticletitle{{Rethinking the Inception Architecture for Computer Vision}}. In \bibinfo{booktitle}{\emph{{IEEE Conference on Computer Vision and Pattern Recognition}}} \emph{(\bibinfo{series}{CVPR})}. \bibinfo{publisher}{IEEE Press}, \bibinfo{address}{Piscataway, NJ, USA}, \bibinfo{pages}{2818--2826}.
\newblock
\urldef\tempurl%
\url{https://doi.org/10.1109/CVPR.2016.308}
\showDOI{\tempurl}


\bibitem[Vaswani et~al\mbox{.}(2017)]%
        {VaswaniShazeer17-attention}
\bibfield{author}{\bibinfo{person}{Ashish Vaswani}, \bibinfo{person}{Noam Shazeer}, \bibinfo{person}{Niki Parmar}, \bibinfo{person}{Jakob Uszkoreit}, \bibinfo{person}{Llion Jones}, \bibinfo{person}{Aidan~N. Gomez}, \bibinfo{person}{Lukasz Kaiser}, {and} \bibinfo{person}{Illia Polosukhin}.} \bibinfo{year}{2017}\natexlab{}.
\newblock \showarticletitle{{Attention Is All You Need}}. In \bibinfo{booktitle}{\emph{{Proceedings of the 31st International Conference on Neural Information Processing Systems}}} (Long Beach, California, USA) \emph{(\bibinfo{series}{NeurIPS})}. \bibinfo{publisher}{Curran Associates, Inc.}, \bibinfo{address}{Red Hook, NY, USA}, \bibinfo{pages}{6000–6010}.
\newblock
\urldef\tempurl%
\url{https://proceedings.neurips.cc/paper/2017/hash/3f5ee243547dee91fbd053c1c4a845aa-Abstract.html}
\showURL{%
\tempurl}


\bibitem[Wang et~al\mbox{.}(2010)]%
        {WangLin2010-gpuKernelFusion}
\bibfield{author}{\bibinfo{person}{Guibin Wang}, \bibinfo{person}{YiSong Lin}, {and} \bibinfo{person}{Wei Yi}.} \bibinfo{year}{2010}\natexlab{}.
\newblock \showarticletitle{{Kernel Fusion: An Effective Method for Better Power Efficiency on Multithreaded GPU}}. In \bibinfo{booktitle}{\emph{{Proceedings of the 2010 IEEE/ACM Int’l Conference on Green Computing and Communications \& Int’l Conference on Cyber, Physical and Social Computing}}} \emph{(\bibinfo{series}{GREENCOM-CPSCOM ’10})}. \bibinfo{publisher}{IEEE Computer Society}, \bibinfo{address}{USA}, \bibinfo{pages}{344–350}.
\newblock
\showISBNx{9780769543314}
\urldef\tempurl%
\url{https://doi.org/10.1109/GreenCom-CPSCom.2010.102}
\showDOI{\tempurl}


\bibitem[Wang et~al\mbox{.}(2022)]%
        {wang2022overlap}
\bibfield{author}{\bibinfo{person}{Shibo Wang}, \bibinfo{person}{Jinliang Wei}, \bibinfo{person}{Amit Sabne}, \bibinfo{person}{Andy Davis}, \bibinfo{person}{Berkin Ilbeyi}, \bibinfo{person}{Blake Hechtman}, \bibinfo{person}{Dehao Chen}, \bibinfo{person}{Karthik~Srinivasa Murthy}, \bibinfo{person}{Marcello Maggioni}, \bibinfo{person}{Qiao Zhang}, \bibinfo{person}{Sameer Kumar}, \bibinfo{person}{Tongfei Guo}, \bibinfo{person}{Yuanzhong Xu}, {and} \bibinfo{person}{Zongwei Zhou}.} \bibinfo{year}{2022}\natexlab{}.
\newblock \showarticletitle{{Overlap Communication with Dependent Computation via Decomposition in Large Deep Learning Models}}. In \bibinfo{booktitle}{\emph{{Proceedings of the 28th ACM International Conference on Architectural Support for Programming Languages and Operating Systems, Volume 1}}} \emph{(\bibinfo{series}{ASPLOS})}. \bibinfo{publisher}{Association for Computing Machinery}, \bibinfo{address}{New York, NY, USA}, \bibinfo{pages}{93--106}.
\newblock
\showISBNx{9781450399159}
\urldef\tempurl%
\url{https://doi.org/10.1145/3567955.3567959}
\showDOI{\tempurl}


\bibitem[Xiong et~al\mbox{.}(2017)]%
        {toward-human-parity-conversational-speech-recognition}
\bibfield{author}{\bibinfo{person}{Wayne Xiong}, \bibinfo{person}{}, \bibinfo{person}{Xuedong Huang}, \bibinfo{person}{Frank Seide}, \bibinfo{person}{}, {and} \bibinfo{person}{Andreas Stolcke}.} \bibinfo{year}{2017}\natexlab{}.
\newblock \showarticletitle{{Toward Human Parity in Conversational Speech Recognition}}.
\newblock \bibinfo{journal}{\emph{{IEEE/ACM Transactions on Audio, Speech, and Language Processing}}}  \bibinfo{volume}{25} (\bibinfo{date}{Sept} \bibinfo{year}{2017}), \bibinfo{pages}{2410--2423}.
\newblock


\end{thebibliography}
